\documentclass{elsart}

\usepackage{amsmath,amssymb}
\usepackage{bm}
\usepackage{graphicx}
\usepackage{subfigure}
\usepackage{verbatim}
\usepackage{wrapfig}
\usepackage[dvips]{color} 
\usepackage{array}

\newcommand \bq{{\mathbf q}}

\def\ln {\mbox{ln}}
\def\tr {\mbox{tr}}

\def\pq2 {((p+\frac{q}{2})^2-M_0^2)((p-\frac{q}{2})^2-M_0^2)}

\def\bq {\mbox{\boldmath{q}}}

\def\slashchar#1{\setbox0=\hbox{$#1$} % set a box for #1
\dimen0=\wd0 % and get its size
\setbox1=\hbox{/} \dimen1=\wd1 % get size of /
\ifdim\dimen0>\dimen1 % #1 is bigger
\rlap{\hbox to \dimen0{\hfil/\hfil}} % so center / in box
#1 % and print #1
\else % / is bigger
\rlap{\hbox to \dimen1{\hfil$#1$\hfil}} % so center #1
/ % and print /
\fi}

\begin{document}

\begin{frontmatter}


\rightline{UT-Komaba/13-14}

\title{Quark-hadron phase transition\\
 in a three flavor PNJL model for interacting quarks}


\author{Kanako Yamazaki and T. Matsui}

\address{Institute of Physics, University of Tokyo \\
Komaba, Tokyo 153-8902, Japan}

\begin{abstract}
% Text of abstract
We extend our previous study of the quark-hadron phase transition at finite temperatures with zero net baryon density by two flavor Nambu-Jona-Lasinio model with Polyakov loop to the three flavor case in a scheme which incorporates flavor nonet 
pseudo scalar and scalar mesonic correlations on equal footing. 
The role of the axial $U(1)$ breaking Kobayashi-Maskawa-'t Hooft interaction on the low-lying thermal excitations is examined. 
At low temperatures, only mesonic correlations, 
mainly due to low mass mesonic collective excitations, pions and kaons, dominate the pressure 
while thermal excitations of quarks are suppressed by the Polyakov loop.  
As temperature increases, kaons and pions melt into 
the continuum of quark and anti-quark excitations successively so that hadronic phase changes
continuously to the quark phase where quark excitations dominate pressure
together with gluon pressure coming from the effective potential for the Polyakov loop.
%We add the gluon pressure in a phenomenological way through the effective
%potential for the Polyakov loop. 
Since we introduce mesons as not elementary fields but
auxiliary fields made from quarks, we can describe the phase transition
between hadronic phase and quark phase in a unified fashion. 

\end{abstract}

%\begin{keyword}
%quark-hadron phase transition


%\end{keyword}

\end{frontmatter}

% main text

%\section{}
%\label{}

\section{Introduction}

This is a sequel to our recent paper\cite{Yamazaki:2012ux}.   
We first briefly summarize the basic motivation of these works before presenting the aim of the present work.

QCD phase transitions in hot and dense matter has been a focus of intense research %one of the most fundamental problems 
in modern nuclear physics\cite{Baym:1984, Pis:1984, Fukushima:2010bq} . 
We all expect that at low temperatures the chiral symmetry is broken spontaneously and all colored objects, including quarks, are confined in color-singlet hadrons, while at high enough temperature the chiral symmetry is restored and the color-confinement is lost, so that the system is composed of a plasma of unconfined nearly massless quarks and gluons, commonly called the quark-gluon plasma.        
The transition of these two limiting states of matter is much less understood, however, due to the difficulty of solving QCD in the non-perturbative regime.  
Although the lattice QCD simulations have been intensively studied for such purposes, there remains a difficulty of extending the method to finite baryon chemical potential due to the sign problem.   
We therefore adopt phenomenological approach using effective models of QCD to describe the 
quark-hadron transition at finite temperature.

The Nambu-Jona-Lasinio (NJL) model has been often used to described the chiral phase transition. 
Although the model was originally formulated\cite{Nambu:1961tp}, before the advent of the quark model, in terms of the hadronic degrees of freedom, it has later been adopted as an effective theory of 
quark dynamics\cite{Hatsuda:1994pi}, respecting chiral symmetry, and has been used to study the QCD phase diagram\cite{Asakawa:1989bq, Fujii:2003bz}. 
A simple way to calculate the equation of state by the model is to perform the mean field approximation. 
In this approximation, the system is composed of thermal excitations of quark quasiparticles even at low temperatures and mesons are treated only as an uniform background field under the mean field theory.
Thermal fluctuations of mesons has been included within the scheme of the NJL model by computing the mesonic correlations\cite{Hufner:1994ma, Zhuang:1994dw, Nikolov:1996jj, Florkowski:1996wf, Oertel:2000jp}. 

Even though the NJL model can be effective to study the chiral phase transition, this model still lacks mechanism of quark confinement. To remedy this problem, the model was extended by Fukushima \cite{Fukushima:2003fw} to include the effect of the Polyakov loop\cite{Polyakov:1978vu, Susskind:1979up} which works as an order parameter of deconfining phase transition. Fukushima's model (PNJL model) has been reformulated as a mean field theory in uniform background temporal color gauge field and has been studied extensively by others\cite{Megias:2004hj, Ratti:2005jh, Ratti:2007jf}. The uniform color gauge field works as imaginary color dependent chemical potential.  Thermal quark excitations are suppressed by the phase cancellations between the distribution functions of triplet of colored quarks at low temperatures where the Polyakov loop is to vanish, while at high temperatures these quark excitations appear as in the original NJL model with vanishing average color gauge field,
  in accord with the Polyakov loop approaching unity.

Mesonic excitations can be found, as in the NJL model, in the mesonic correlations beyond the mean field approximation \cite{Yamazaki:2012ux, Rossner:2007ik, Blaschke:2007np, Hansen:2006ee,  Wergieluk:2012gd, Benic:2013tga}. We have shown in \cite{Yamazaki:2012ux} that equation of state of a meson gas can be derived explicitly at low temperatures by the method of auxiliary fields which physically express effective meson fields build-up as a quark-anti-quark bound states as in the original Nambu-Jona-Lasinio model.
The purpose of this work is to extend the previous study based on the two flavor PNJL model to three flavor model.

Even though the analysis with two flavor model gives us rich physics, once we hope to compare our results with experimental data, it is necessary to consider the role of the strangeness degree of freedom. In this paper, we extend our previous work with two flavor model\cite{Yamazaki:2012ux}  to the three flavor case along the line of the NJL model \cite{Kunihiro:1987bb, Bernard:1987sg, Klimt:1990ws, Klevansky:1992qe, Buballa:2003qv}. 
By changing quark fields with two flavor components to three flavors, the number of mesons which appear in this model also changes. 
Unlike the two flavor model which contains three pions and sigma meson, there are nine pseudo scalar mesons (3$\pi $, 4$K$, $\eta$ and $\eta'$) and nine scalar mesons ($\sigma$, 4$\kappa $, $f_0$ and 3$a_0$) in the three flavor model.  As for the scalar mesons, all scalar mesons have not been established by experimental data\cite{Beringer:1900zz}
due to their large decay widths. 
However some analyses support for the existence of the scalar nonet \cite{Ishida:1999qk, Ishida:1997wn, Fariborz:2009cq}. In addition to the change of quark fields, it is necessary to take the six point interaction called the Kobayashi-Masukawa-'t Hooft interaction\cite{Kobayashi:1970ji, 'tHooft:1976up}. This interaction breaks axial U(1) symmetry, ensuring the observed mass splitting of $\eta $ and $\eta '$.

Extension of the flavor number has been performed also in the PNJL model \cite{Abuki:2008ht, Fukushima:2008wg, Powell:2011ig, Kouno:2013zr}. However, all of these works have been done under the mean field approximation, so that mesonic excitations are absent in these calculations.  
In this work we put mesonic correlations into the equation of state and we describe how the degrees of freedom of thermal excitations change from those of hadrons to quark and gluons.  

%%%%%%
The rest of this paper is organized as follows. 
In the next section, we introduce a three flavor PNJL model to be used in the evaluation of the path integral expression for the partition function. 
All NJL type models contain four-point interactions of fermion fields. 
These four point fermionic interactions can be eliminated by standard Hubbard-Stratonovic transformation in favor of integrable quadratic terms.. 
Three flavor NJL models has additional six-point interaction, however. 
To eliminate this six-point interaction we need to introduce "counter terms" generated by the third power of bosonic auxiliary fields each shifted by the quark bilinear terms with appropriate normalization.   
Intuitively, this procedure may be regarded as reducing the six-point interaction to effective four-point interactions by replacing one set of quark bilinear term by their expectation value\cite{Klimt:1990ws}.  
In Section 3, we summarize the results of the mean field approximation which freezes meson fields as back ground fields. 
We also show how the effect of the Polyakov loop appears in the equation of state. 
In Section 4, we calculate the contribution of mesonic correlations to the equation of state and show that pressure are dominated by low mass mesons as pseudo-Nambu-Goldstone modes, pions and kaons,  at low temperatures, while it is dominated by quarks and gluons at high temperatures. In order to explore what is happening at intermediate temperatures, we also calculate up to which temperature collective mesonic excitation persist. 
We summarize this work in Section 5.

\section{Model setup}

In this section, we set a three flavor PNJL model and derive thermodynamic potential by calculating a partition function by the path integral method. We introduce the Lagrangian of a three flavor PNJL model:
\begin{eqnarray}
\mathcal{L}=\sum_{i,j=1}^3\bar{q}_i(i\slashchar{D}-\hat{m})_{ij}q_j+\mathcal{L}_4+\mathcal{L}_6
\end{eqnarray}
where
\begin{eqnarray}
\mathcal{L}_4=G\sum_{a=0}^8\bigl[ (\bar{q}\lambda^aq)^2+(\bar{q}i\gamma_5\lambda^aq)^2\bigr] 
\end{eqnarray}
and
\begin{eqnarray}
\mathcal{L}_6&=&-K\bigl[ \mbox{det}\ \bar{q}(1+\gamma_5)q+\mbox{det}\ \bar{q}(1-\gamma_5)q\bigr] 
\end{eqnarray}
for three flavor light quarks, $\bar{q}=(\bar{q}_1, \bar{q}_2, \bar{q}_3)=(\bar{u}, \bar{d}, \bar{s})$.  $D_{\mu }=\partial _{\mu }+gA_0\delta _{\mu , 0}$ where $A_0$ is the temporal component of gauge fields, 
$A_0=-iA_4$\cite{KG06}.  
Gauge field is not treated here as a dynamical valuable but as an external parameter like imaginary chemical potential which depends on the color of quarks. $\hat{m}$ is a $3\times 3$ mass matrix, giving bare quark masses $m_u$, $m_d$ and ms for u, d and s quarks, respectively. In the later calculations, we set $m_u = m_d = m$ assuming isospin symmetry.

\begin{figure}[htbp]
\begin{center}
\includegraphics[clip,width=70mm]{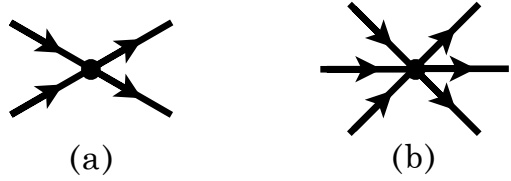}
\end{center}
\caption{
(a) 4 point interaction, (b) 6 point interaction
}
\label{fig:interactions}
\end{figure}

$\mathcal{L}_4$ is  a four-point interaction of quark and anti-quark (Fig.\ref{fig:interactions}, (a)), with G the coupling strength. Each $\lambda^a$ is a $3\times 3$ matrix in flavor space, with $a$ running from 0 to 8. 
$\lambda^1$ to $\lambda^8$ are the Gell-Mann matrices and $\lambda^0$ is proportional to an identity matrix: $\sqrt{\mathstrut 2/3} I$. 
The relative strength of scalar and pseudo scalar interactions is determined so that they as a whole possess the chiral symmetry.
$\mathcal{L}_6$ is a six-point interaction, called the Kobayashi-Maskawa-'t Hooft interaction (Fig.\ref{fig:interactions}, (b)). with $K$ the strength of the interaction. Since $q$ and $\bar{q}$ each have three components in flavor $SU(3)$, $\mathcal{L}_6$ consists of 6th power terms of fermion fields. $\mathcal{L}_6$ can be written in the following form\cite{Vogl:1991qt, Klimt:1989pm}:
\begin{eqnarray}
\mathcal{L}_6= \frac{K}{6}d_{abc}\Bigl[ \frac{1}{3}(\bar{q}\lambda^iq)(\bar{q}\lambda^jq)
+(\bar{q}\gamma_5\lambda^iq)(\bar{q}\gamma_5\lambda^jq)\Bigr] (\bar{q}\lambda^kq) .
\end{eqnarray}
where $d_{abc}$ contain symmetric constants for $SU(3)$ for $a = 1, \cdots, 8$, besides $d_{000} = \sqrt{2/3}$, $d_{0bc}=-\sqrt{1/6}$ ($b, c \neq 0$).

The partition function is given by
\begin{eqnarray}
Z(T, A_4)=\int [dq][d\bar{q} ]\mbox{exp}\bigl[ \int_0^{\beta }d\tau \int d^3x \mathcal{L} (q, \bar{q}, A_4)\bigr]  .
\end{eqnarray}
From the interaction terms, this model can incorporate correlations which generate pseudo scalar mesons and scalar mesons. For pseudo scalar mesons, there are nine mesons, three kinds of $\pi$, four kinds of $K$, $\eta$ and $\eta'$. They make a nonet in SU(3) flavor classification.  In the chiral limit,  mass of all mesons is exactly zero, in this limit, forming the nine massless Nambu- Goldstone modes. The axial U(1) symmetry is broken by the 6-point interaction, making $\eta^0$ massive. In addition, the SU(3) flavor symmetry is broken due to the non-vanishing bare quark masses, $m_u$, $m_d$, $m_s$, generating the physical masses of each meson.

There appear also nine scalar mesons in this scheme, not all of which are confirmed by experiments. Especially the existence of $\kappa $ is still very controversial. Beside $\kappa $, all other scalar mesons are listed on the data compiled by the particle data group\cite{Beringer:1900zz}.

The original model Lagrangian contains 4th and 6th power of fermion fields. 
These non-quadratic terms make difficult to perform the fermion integrals in the partition function. 
Recalling the two flavor case, the PNJL model has only four point interactions which we eliminate by generating "counter terms" contained in the square of the auxiliary bosonic field shifted by the quadratic quark fields.
Integration of the quark fields can be performed analytically and the partition function is written in terms the path integral over the bosonic fields. 
This standard Hubbard-Stratonovic transformation\cite{Hu59, St57} cannot be applied directly in the presence of the six point fermionic  interaction.
To eliminate the six point interactions we need to introduce extra "counter terms" generated by the third power of the 
auxiliary bosonic fields shifted by a bilinear form of the quark fields with appropriate normalization, reducing the six point interaction to effective four point interactions.  
This procedure shift the coupling $G$ of the 4th order quark fields which can be eliminated by the standard procedure.
%by inserting dummy integrals over auxiliary bosonic fields.  

To be more explicit, we introduce auxiliary bosonic fields $\phi ^a$ and $\pi ^a$ coupled to quark scalar densities $\bar{q}\lambda ^a q$ and 
pseudo scalar density $i\bar{q}\gamma _5\lambda^a q$ respectively by multiplying $Z(T, A_4)$ by a constant dummy integral:
\begin{eqnarray}
\int [d\phi ][d\pi ]\mbox{exp}\bigl( \int _0^{\beta }d\tau \int d^3 x \mathcal{L}_{b} ( \phi, \pi ) \bigr)
%&& \frac{K}{18}d_{abc}  (\phi ^a+\bar{q}\lambda ^aq) \bigl(  (\phi ^b+\bar{q}\lambda ^bq)  (\phi ^c+\bar{q}\lambda ^cq) 
 %\nonumber \\
% && \qquad + (\pi ^b+i\bar{q}\lambda ^b \gamma_5 q )(\pi ^c+i\bar{q}\lambda ^c \gamma_5 q )   \bigr) \Bigr]   ??
\end{eqnarray}
with
\begin{eqnarray}
\mathcal{L}_{b} & = & \frac{K}{24G^3}d_{abc}  (\phi^a -2G \bar{q}\lambda^a q ) \bigl[ \frac{1}{3}(\phi^b -2G \bar{q}\lambda^b q )  ( \phi^c-2G\bar{q}\lambda^c q ) \nonumber \\
& &    
 + (\pi^b -2G i\bar{q}\lambda^b \gamma_5 q  )(\pi^c-2G i\bar{q}\lambda^c \gamma_5 q )   \bigr] 
 \nonumber \\
& & \qquad   
 + \frac{K}{24G^2}d_{abc} \phi^a  \bigl[ ( \phi^b -2G \bar{q}\lambda^b q )  (\phi^c-2G\bar{q}\lambda^c q ) \nonumber \\
 & & \qquad 
 +  (\pi^b-2G i\bar{q}\lambda^b \gamma_5 q  )(\pi^c-2G i\bar{q}\lambda^c \gamma_5 q )   \bigr] 
  \nonumber \\
& & \qquad   \qquad 
- \frac{1}{4G} \bigl[ (\phi^a-2G \bar{q}\lambda^aq)^2+ (\pi^a-2G i\bar{q}\lambda ^a \gamma_5 q )^2  \bigr] 
\end{eqnarray}
Only second power of the pseudo-scalar fields can appear in order to respect the Lorentz symmetry of the Lagrangian.  
The desired "counter terms" for the 6 point quark interactions can be found in the expansion of the first term, 
which however also generates extra 4 point quark interactions; they are removed by the "counter term" generated by the second term. 
The third term is to eliminate the 4 point quark interactions in the original Lagrangian. 
%This procedure corresponds in the mean field approximation to replacing a pair of quark and anti-quark by a condensate (Fig.\ref{fig:eff_4p}).   

Adding $\mathcal{L}_{b}$, the original Lagrangian is converted to a form which contains the quark fields only in the bilinear form
in addition to the second and third power terms of the auxiliary bosonic fields: 
\begin{eqnarray}
\mathcal{L} + \mathcal{L}_{b} & = & \sum_{i,j=1}^3\bar{q}_i(i\slashchar{D}-\hat{m} - \Sigma (\phi_a, \pi_a) )_{ij}q_j 
+ \frac{K}{18G^3}d_{abc} \phi^a \phi^b \phi^c \nonumber \\ 
& & \qquad \qquad  - \frac{K}{6G^3}d_{abc} \phi^a \pi^b \pi^c - \frac{1}{4G} ( \phi_a^2 + \pi_a^2 )
\end{eqnarray}
where
\begin{equation}
\Sigma (\phi_a, \pi_a) = \frac{K}{4 G^2} d_{abc} \lambda^a \phi^b \phi^c  %+ i  \gamma_5 \pi^c \bigr]  
-   \lambda^a \bigl[  \phi^a + i \gamma_5 \pi^a \bigr]
\end{equation}
is the self-energy matrix of quark quasiparticles due to the coupling to the auxiliary fields. 
%This would generate flavor mixing.

Now the integration over the Grassmann quark fields can be performed and we obtain the effective action written in terms of the auxiliary bosonic fields $\phi_a$ and $\pi_a$:
\begin{eqnarray}
Z(T, A_4)=\int [d\phi][d\pi ] e^ 
%\mbox{exp}\bigl[ 
{ - I  (\phi, \pi, A_4) } %\bigr]  .
\end{eqnarray}
where
\begin{eqnarray}
 I  (\phi, \pi, A_4) & = &\frac{1}{\beta V}  \int_0^{\beta} d\tau \int d^3x 
 \bigl[ {\rm tr}_q ~ \ln ( \beta S_E^{-1} ) + \frac{K}{18G^3}d_{abc} \phi^a \phi^b \phi^c 
 \nonumber \\
&& \qquad  
 - \frac{K}{6G^3}d_{abc} \phi^a \pi^b \pi^c - \frac{1}{4G} ( \phi_a^2 + \pi_a^2 ) \bigr]
\end{eqnarray}
with the "inverse Euclidean quark propagator" given by
\begin{equation}
S_E^{-1} =   i \slashchar{D}_E + \hat{m} + \Sigma (\phi, \pi, A_4) 
\end{equation}
with $\slashchar{D}_E = \sum_{i = 1, \cdots, 3} \gamma_i \partial_i +  \gamma_4 (\partial_\tau +i g A_4)$. 
The trace $\mbox{tr}_q$ includes sum over the color, and the Dirac spinor indices of the quark fields.  

In order to calculate the pressure of mesonic correlation, we expand an effective action up to the second order of fluctuations around a stationary point,  $\varphi_a = \phi_a -  {\bar \phi}_a$, 
%which is chosen to be characterized by non-vanishing $\phi_0$,  while 
%other auxiliary fields are set zero,% $\phi_1 = \phi_2 = \cdots = \phi_8 =0$ and $\pi_0 = \cdots = \pi_8 = 0$. 
\begin{eqnarray}
I(\phi , \pi , A_4)= I_0+\frac{1}{2}\left. \frac{\delta ^2 I}{\delta \phi _a\delta \phi _b}\right|_{\phi = {\bar \phi}} \varphi_a 
\varphi_b \ +\left. \frac{1}{2}\frac{\delta ^2 I}{\delta \pi _a\delta \pi _b}\right|_{\phi =\phi _0} \pi_a \pi_b \ \cdots 
\label{eff_act}
\end{eqnarray} 
with
\begin{equation}
I_0 = I( \bar{\phi},  \pi = 0 , A_4)
\end{equation}
where the stationary value of $\bar{\phi}_a$ is determined by the condition:
\begin{eqnarray}
\left. \frac{\delta I}{\delta \phi_a }\right|_{\phi =\bar{\phi}}=0 .
\label{SC}
\end{eqnarray}
We have assumed that the stationary values of the pseudo scalar fields $\pi_a$ all vanish.
Keeping only up to quadratic terms in expansion, 
\begin{eqnarray}
Z(T , A_4)\simeq e^{- I_0}\int [d\phi ][d\pi ]\mbox{exp}\biggl[ -\frac{1}{2}\left. \frac{\delta ^2 I}{\delta \phi _a\delta \phi _b}\right|_{\phi =\bar{\phi}} \varphi_a \varphi_b -\left. \frac{1}{2}\frac{\delta ^2 I}{\delta \pi _a\delta \pi _b}\right|_{\phi =\bar{\phi}}\pi_a \pi_b  \biggr] .\ \ \ \ \ 
\nonumber \\
\end{eqnarray}
If we stop the expansion at the second order fluctuation, ignoring interactions of mesons, we can perform the Gaussian integral over the meson fields. Then we get the thermodynamic potential;
\begin{eqnarray}
\Omega (T , A_4)=T \ln Z = -T\Bigl( I_0+\frac{1}{2}\mbox{Tr}_M\mbox{ln}\frac{\delta ^2I}{\delta \phi_a\delta \phi_b} +\frac{1}{2}\mbox{Tr}_M\mbox{ln}\frac{\delta ^2I}{\delta \pi_a\delta \pi_b}\Bigr) . \label{TP}
\end{eqnarray}
The first term of Eq.(\ref{TP}) represents the thermodynamic potential under the mean field approximation and the second and third terms represent the contribution of mesonic correlations to the thermodynamic potential.

\section{Mean field approximation}  

The thermodynamic potential in the mean field approximation, $\Omega _{MF}(T , A_4)$ and the corresponding pressure $p_{MF}(T, A_4)$ have a relation to the leading term of the effective action Eq.(\ref{eff_act}), $I_0$ :
\begin{eqnarray}
\Omega _{MF}(T, A_4)= - TI_0=-p_{MF}(T, A_4)V 
\end{eqnarray}
The explicit form of the leading term $I_0$ is given by 
\begin{eqnarray}
I_0&=&  \beta V \Bigl[ - \frac{1}{4G} ({\bar \phi}_u^2 +{\bar \phi}_d^2+{\bar \phi}_s^2)
+\frac{K}{2G^3}{\bar \phi}_u {\bar \phi}_d {\bar \phi}_s\nonumber \\
&+& 2T\sum_i \sum_n \int \frac{d^3p}{(2\pi )^3}\mbox{tr}_c \mbox{ln}\bigl[ \beta ^2 \bigl( (\epsilon_n-gA_4)^2+\mathbf{p}^2+M_i^2\bigr) \bigr] \Bigr]  \label{I_0}
\end{eqnarray}
where $\epsilon_n$ is fermionic Matsubara frequencies, $\epsilon_n=(2n+1)\pi T$ and the trace is to be performed over the $3\times 3$ color matrix $A_4$.
We have introduced for convenience the notations:
\begin{eqnarray}
{\bar \phi}_u & \equiv & 2G\langle \bar{u}u\rangle =\frac{2}{\sqrt{6}}{\bar \phi}_0+{\bar \phi}_3 +\frac{2}{2\sqrt{3}}{\bar \phi}_8 \label{phiu}\\ 
{\bar \phi}_d & \equiv & 2G\langle \bar{d}d\rangle =\frac{2}{\sqrt{6}}{\bar \phi}_0 -{\bar \phi}_3+\frac{2}{2\sqrt{3}}{\bar \phi}_8 \\
{\bar \phi}_s & \equiv & 2G\langle \bar{s}s\rangle =\frac{2}{\sqrt{6}}{\bar \phi}_0-\frac{2}{\sqrt{3}}{\bar \phi}_8 \label{phis}
\end{eqnarray}
as implied by the relation ${\bar \phi}_a = G \langle \bar{q} \lambda_a q\rangle$ in the mean field approximation.
The constituent quark masses $M_i$ in Eq.(\ref{I_0}) are given in terms of $\langle \bar{q_i}q_i\rangle $ 
defined by Eqs.(\ref{phiu})-(\ref{phis});
\begin{eqnarray}
M_u&=&m_u-4G\langle \bar{u}u\rangle +2K\langle \bar{d}d\rangle  \langle \bar{s}s\rangle  \label{Gapu}\\
M_d&=&m_d-4G\langle \bar{d}d\rangle +2K\langle \bar{s}s\rangle  \langle \bar{u}u\rangle  \label{Gapd}\\
M_s&=&m_s-4G\langle \bar{s}s\rangle +2K\langle \bar{u}u\rangle  \langle \bar{d}d\rangle \label{Gaps} .
\end{eqnarray}
Since $\langle \bar{q}_i q_i \rangle $ is related to the Euclidean $i$-quark propagator $S^i_E  =  ( i \slashchar{D}_E + M_i )^{-1}$ by  
\begin{eqnarray}
\langle \bar{q}_i q_i \rangle = -i\mbox{Tr}S_E^i = T \sum_n \int \frac{d^3 p}{(2\pi)^3} \frac{M_i}{(\epsilon_n - g A_4)^2 + \mathbf{p}^2 + M_i^2} ,
\end{eqnarray}
where the fermionic Matsubara frequency sum can be evaluated by the method of contour integration\cite{FW71}
\begin{eqnarray}
\langle \bar{q}_i q_i \rangle 
& = & \int \frac{d^3 p}{(2\pi)^3} \frac{M_i}{E_i (p)} \left[ - 1 + 2 f (E_i (p) - i gA_4) \right] 
\end{eqnarray}
with $E_i (p) = \sqrt{\mathbf{p}^2 + M_i^2}$ and $f (E) = 1/(e^{\beta E} +1)$. 

Eqs.(\ref{Gapu})-(\ref{Gaps}) are also written as 
\begin{eqnarray}
M_i = m_i+4iG\mbox{tr}S_E^i-2K \epsilon_{ijk} (\mbox{tr}S_E^j)(\mbox{tr}S_E^k) .
\end{eqnarray}
which are equivalent to the stationary conditions to the auxiliary scalar fields Eq.(\ref{SC}).
These equations are a three flavor extension of the Nambu-Jona-Lasinio 
gap equation which determines the quark masses $M_i$ (gaps in the single particle energy spectra) self-consistently.

In the following calculation, we assume unbroken isospin symmetry so that $u$-quark and $d$-quark are degenerate.  
We then find the pressure in the mean field approximation for the three flavor model,
\begin{eqnarray}
p_{MF}(T, A_4)=&-&\frac{1}{4G} (2{\bar \phi}_u^2+{\bar \phi}_s^2)
+\frac{K}{2G^3}{\bar \phi}_u^2{\bar \phi}_s\nonumber \\
&+&2T\sum_i \sum_n \int \frac{d^3p}{(2\pi )^3}\mbox{tr}_c \mbox{ln}\bigl[ \beta ^2 \bigl( (\epsilon_n-gA_4)^2+\mathbf{p}^2+M_i^2\bigr) \bigr] 
\end{eqnarray}
%where $\epsilon_n$ is fermionic Matsubara frequencies, $\epsilon_n=(2n+1)\pi T$, the trace is to be performed over the $3\times 3$ color matrix $A_4$.  
%The sum is taken over the Matsubara frequencies. 
Evaluating the discrete sum over the Matsubara frequencies by the standard method of contour integration, we find
\begin{eqnarray}
p_{MF}(T, A_4)=p_{MF}^0
+2 \sum_i \int \frac{d^3p}{(2\pi )^3}\frac{\mathbf{p}^2}{3E_i }\mbox{tr}_c 
\bigl[ f(E_i+igA_4)+f(E_i-igA_4)\bigr] \label{p_mf}\ \ \ \ \ \ 
\end{eqnarray}
where 
\begin{eqnarray}
p_{MF}^0=3\times 2\sum_i \int^\Lambda \frac{d^3p}{(2\pi )^3}E_i (p)
-\frac{1}{4G}(2{\bar \phi}_u^2+{\bar \phi}_s^2) +\frac{K}{2G^3}{\bar \phi}_u^2 {\bar \phi}_s 
\end{eqnarray}
is the pressure exerted by the quark condensate and the zero point motion of the quark quasiparticles with energy 
$E_i=\sqrt{\mathbf{p}^2+M_i^2}$, $i=u, d, s$. 
\begin{eqnarray}
f(E_i\pm igA_4)=\frac{1}{e^{\beta (E_i\pm igA_4)}+1}
\end{eqnarray}
is the quark (anti-quark) quasiparticle distribution function in the external gauge field.
%The constituent quark mass $M_i$ is determined by gap equations obtained from the stationary conditions (\ref{SC}) for the auxiliary scalar fields.

In the above expressions, the constant temporal gauge field $A_4$ appears as a phase factor together with the quark quasiparticle energy in the quark (anti-quark) distribution function. It looks very similar to the gauge invariant (path-ordered) Polyakov loop phase integral,
\begin{eqnarray}
L(\mathbf{r})=\mathcal{P}\mbox{exp}\Bigl[ ig\int _0^\beta d\tau A_4(\mathbf{r}, \tau )\Bigr] 
\end{eqnarray}
whose thermal expectation value measures the extra free energy associated with the color charge in fundamental representation fixed at a spatial point $\mathbf{r}$. Although this connection is not strict, since quarks are moving in a uniform background gauge field not fluctuating either in space or in imaginary time to replace the phase factor in the quark quasiparticle distribution function by the thermal average of the Polyakov loop.
We replace $\langle \frac{1}{3}\mbox{tr}_c f(E_i+igA_4)\rangle $ by 
\begin{eqnarray}
f_{\Phi }(E_i)=\frac{\bar{\Phi }e^{2\beta E_i}+2\Phi e^{\beta E_i}+1}{e^{3\beta E_i}+3\bar{\Phi }e^{2\beta E_i}+3\Phi e^{\beta E_i}+1}
\end{eqnarray}
where $\Phi =\frac{1}{3}\langle \mbox{tr}_c L\rangle $ 
and $\bar{\Phi }=\frac{1}{3}\langle \mbox{tr}_c L^{\dagger }\rangle $. 
$\Phi $ and $\bar{\Phi }$ behave as an order parameter of de-confining phase transition.    

We apply the same procedure to the quark condensates which appears in the gap equation 
so that $\langle \bar{q}_i q_i \rangle$ is replaced by 
\begin{eqnarray}
\langle \langle \bar{q}_i q_i \rangle \rangle
& = & - \int_\Lambda \frac{d^3 p}{(2\pi)^3} \frac{M_i}{E_i (p)} 
 +  2  \int \frac{d^3 p}{(2\pi)^3} \frac{M_i}{E_i (p)} f_\Phi (E_i (p) )
\end{eqnarray}
where we indicated a momentum cut-off at $p = \Lambda$ in the otherwise divergent vacuum polarization term. 

Here we determine $\Phi $ phenomenologically by adding an effective potential of $\Phi $:
\begin{eqnarray}
\Omega(T, \Phi)=\langle \Omega (T, A_4)\rangle +\mathcal{U}(T, \Phi )
\end{eqnarray}
where we define, following,
\begin{eqnarray}
\mathcal{U}(T, \Phi )/T^4=-\frac{1}{2}b_2(T)\bar{\Phi}\Phi-\frac{1}{6}b_3(\bar{\Phi}^3+\Phi^3)+\frac{1}{4}b_4(\bar{\Phi}\Phi)^2 
\end{eqnarray}
with 
\begin{eqnarray}
b_2(T)=a_0+a_1\Bigl( \frac{T_0}{T}\Bigr) +a_2\Bigl( \frac{T_0}{T}\Bigr) ^2+a_3\Bigl( \frac{T_0}{T}\Bigr) ^3 . 
\end{eqnarray}
The parameters are chosen so that $\Phi =0$ at low temperatures, $\Phi $ gets close to $1$ at high temperatures and $\mathcal{U}(T, \Phi)$ gives pressure of gluons obtained by lattice calculations at high temperatures.  

After replacing the quark distribution function by the statistical average over the gauge field $A_4$, the pressure under mean field approximation is given by
\begin{eqnarray}
p_{MF}(T)=p_{MF}^0+2\times 3\sum_i \frac{d^3p}{(2\pi )^3} \frac{\mathbf{p}^2}{3E_i} f_{\Phi }(E_i)
-\mathcal{U} (T, \Phi )
\end{eqnarray}

\begin{figure}[htbp]
\begin{center}
\includegraphics[clip,width=80mm, angle=270]{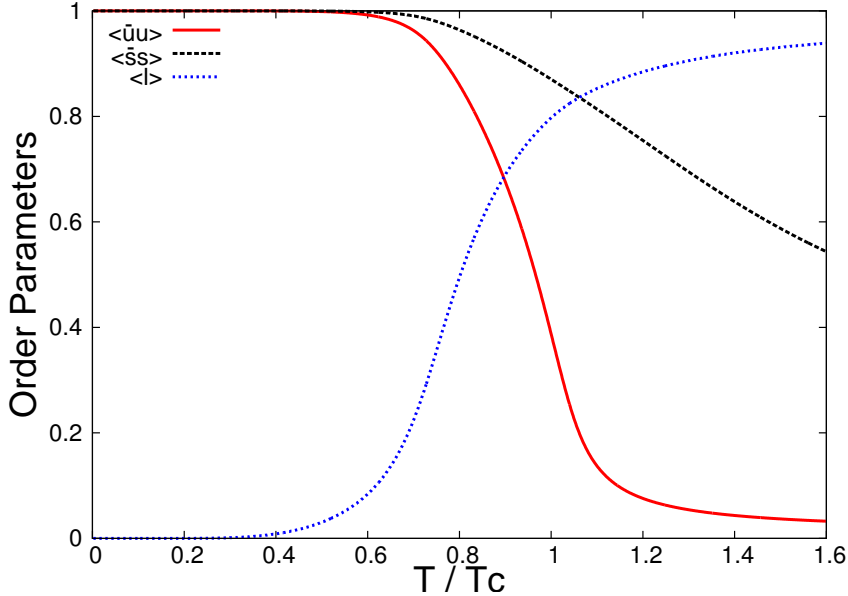}
\end{center}
\caption{
Temperature dependence of the order parameters in the mean field approximation. The bare quark mass is taken 5.5 MeV for $u$ and $d$ quarks, and 140.7 MeV for $s$ quark.  The chiral condensates are normalized by their vacuum expectation values: 
$\langle \bar{u}u\rangle^{1/3} = - 241.9$ MeV, $\langle \bar{s}s\rangle^{1/3} = - 257.7$ MeV.
}
\label{fig:gapeq}
\end{figure}

We show in Fig.\ref{fig:gapeq} temperature dependence of the order parameters, $\langle \bar{u}u\rangle$,  $\langle \bar{s}s\rangle$  the amplitudes of the chiral condensates and $\langle l\rangle =\Phi$ the expectation value of the Polyakov loop.
Since both the chiral and the confining transitions become crossover in this calculation,  the temperature is scaled by $T_c$ the pseudo critical temperature determined by the maximum of the chiral susceptibility,
the second derivative of the pressure with respect to $\langle \bar{u}u\rangle$.  
In this calculation, $T_c$ is found $220\mbox{MeV}$.  

\begin{table}[htb]
  \begin{center}
    \caption{Parameters}
    \begin{tabular}{|c|c|c|c|c|} \hline
    $m_u=m_d$ & $M_s$ & $\Lambda $ & $G\Lambda^2$ & $K\Lambda^5$ \\ \hline \hline
      $5.5$ MeV & $140.7$ MeV & $602.3$ MeV & $1.835$ & $12.36$ \\ \hline
      \end{tabular}
  \end{center}
\end{table}

We choose the values of the parameters in accordance with \cite{Rehberg:1995kh}; 
 $m_u = m_d = 5.5$MeV, $m_s = 140.7$MeV, $\Lambda =602.3$MeV, $G\Lambda ^2=1.835$ and $K\Lambda^5=12.36$. 
These parameters are determined by the pion mass $m_{\pi} =135.0$MeV, the kaon mass $m_{K}=497.7$MeV, the $\eta '$ mass $957.8$MeV and the pion decay constant $f_{\pi}=92.4$MeV in vacuum.
%These values of these parameters are determined at $T = \mu =0$. 
Note that the effect of the Polyakov loop enters only through the quark distribution function so that it does not appear in vacuum.
This implies that our procedure to set the values of the parameters by the physical observable in the vacuum is the same as the one 
taken for the NJL model without confinement. 
The solid red line and the dotted black line are the results of two chiral condensates, $\langle \bar{u}u\rangle$,  $\langle \bar{s}s\rangle$ respectively,  scaled by the vacuum expectation value of each condensate.   
The dotted blue line is expectation value of the Polyakov loop which characterizes the deconfining transition.
The amplitude of the $u$-quark condensate approaches zero rapidly at temperatures above $T_c$, while that of the $s$-quark condensate remains non-zero even at higher temperature due to the larger bare $s$-quark mass, which is comparable to $T_c$.  

\begin{figure}[htbp]
\begin{center}
\includegraphics[clip,width=80mm, angle=270]{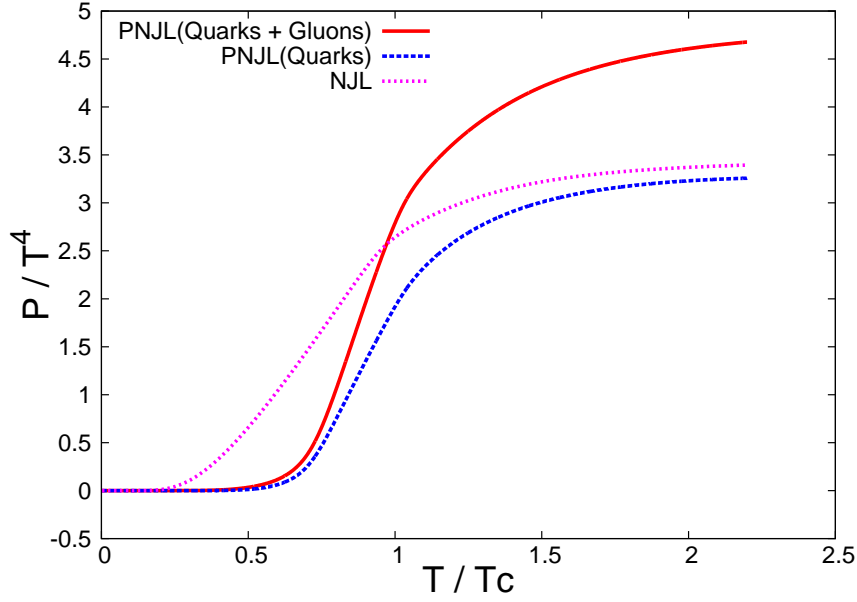}
\end{center}
\caption{
Pressures in the mean field approximation.  The dotted blue line is the pressure of quarks calculated with the PNJL model,
the solid red line includes the gluon pressure coming from the effective potential for the Polyakov loop.  The dotted pink line is the quark pressure calculated with the NJL without confinement, which allows quark excitations even at low temperatures. 
In the PNJL model both quark and gluon pressure deplete rapidly below crossover region due to the Polyakov loop which suppresses quark distribution in the confining phase. 
}
\label{fig:PMF}
\end{figure}

We plot the pressure in the mean field approximation in Fig.\ref{fig:PMF}. 
The solid red line (the dotted blue line) is calculated in the mean field approximation with (without) the
effective potential of the Polyakov loop $\mathcal{U}(T)$, which contains the gluon pressure; the dotted blue line is the pressure only due to the quark quasiparticles. 
The dotted pink line is the pressure calculated by the NJL model. 
Comparing the dotted red, pink and blue lines, one sees that the quark pressure becomes almost zero at low temperatures because the quark excitations are strongly suppressed by the Polyakov loop in the confining phase, 
while they persist even at low temperatures in the NJL model without confinement.  

\section{Mesonic correlation}

In the previous section, we have discussed the equation of state obtained by the mean field approximation with the three flavor PNJL model.  
It is seen that the Polyakov loop depletes the quark pressure in the low temperature confining phase. 
Now we explore how mesonic correlations contribute to the EOS in this section.

The pressure from mesonic correlations can be calculated from the second and the third terms of Eq.(\ref{TP}) for
the thermodynamic potential.  
By the thermodynamic relation $ p V = - T \Omega $ ,  the pressure of mesonic correlations in the background gauge field $A_4$ is given by 
\begin{eqnarray}
p_M(T, A_4) = -\frac{T}{2V}\bigl( \mbox{Tr}_M \mbox{ln}\frac{\delta ^2I}{\delta \phi_a\delta \phi_b} +\frac{1}{2}\mbox{Tr}_M\mbox{ln}\frac{\delta ^2I}{\delta \pi_a\delta \pi_b}\bigr) \label{pm}
\end{eqnarray}
where $I$ is an effective action, Eq.(\ref{eff_act}). 
The first term is a contribution from scalar mesons and the second term is from pseudo scalar mesons. 
The indices $a$ and $b$ in Eq.(\ref{pm}) run 0 to 8 in the SU(3) flavor space.
The trace $\mbox{Tr}_M$ is taken over the space-time coordinates of the auxiliary meson fields which obey a periodic boundary condition in the imaginary time direction.  

From Eqs.(\ref{eff_act}) and (\ref{pm}), we find
\begin{eqnarray}
p_{M} & = & -\sum _n \int \frac{d^3q}{(2\pi )^3} \Bigl\{ 3\mbox{ln} \mathcal{M}_{\pi }(\omega _n, q)+4\mbox{ln}\mathcal{M}_{K}(\omega _n, q) +\mbox{ln}\mathcal{M}_{\eta }(\omega _n, q) \nonumber \\
& & \qquad \qquad +\mbox{ln}\mathcal{M}_{\eta ' }(\omega _n, q) 
+ \mbox{ln}\mathcal{M}_{\sigma }(\omega _n, q)+4\mbox{ln}\mathcal{M}_{\kappa }(\omega _n, q)
\nonumber \\
& & \qquad \qquad \qquad + 3\mbox{ln}\mathcal{M}_{a_0}(\omega _n, q)+\mbox{ln}\mathcal{M}_{f_0 }(\omega _n, q)\Bigr\} \label{p_mesons} \nonumber \\
\label{mpressure}
\end{eqnarray}
where $\mathcal{M}_\alpha$ measures the Gaussian fluctuation of the Fourier component of the mesonic auxiliary fields in the mesonic channel $\alpha ( = \pi, K, \eta, \eta', \sigma, \kappa, a_0, f_0 )$ 
with the Matsubara frequency $\omega_n = 2n \pi T$ and the spatial inverse wavelength $q$; it contains the information about the existence of collective mesonic excitations in each channel.
It is given in the form
\begin{eqnarray}
\mathcal{M}_\alpha (\omega_n, q)= \frac{1}{2G'_\alpha}-\Pi _\alpha (\omega_n, q) , 
%\quad \hbox{for} \quad M = \pi, K, \eta, \eta', \sigma, \kappa, a_0, f_0
\label{M}
\end{eqnarray}
where $\Pi _\alpha (\omega_n, q)$ is the quark polarization for the mesonic channel $\alpha$, %which can be decomposed into two parts: 
\begin{eqnarray}
\Pi_\alpha (\omega_n, q) & =  & \beta \sum_m \int \frac{d^3 p}{(2 \pi)^3} \langle \tr_q 
\bigl( \Lambda_\alpha S_E ( \omega_n + \epsilon_m, \mathbf{p} +  \mathbf{q}, A_4) \Lambda_\alpha S_E (\epsilon_m, \mathbf{p}, A_4) \bigr) \rangle
\nonumber \\
%& = & \Pi_M^1(A_4) + \Pi_M^2(\omega_n,q, A_4). \label{Pi}
%\nonumber \\
\label{PiS}
\end{eqnarray}
%where $\Pi_M^1 ( A_4)$ is a non-dispersive contact term which may be written as
%\begin{equation}
%\Pi_M^1(A_4)  =  T \sum_n \int \frac{d ^3 p}{(2\pi)^3} 
%\end{equation}
for a scalar field of the channel $\alpha = \sigma, \kappa, a_0, f_0$ and 
\begin{eqnarray}
\Pi_\alpha (\omega_n, q) =  \beta  \sum_m \int \frac{d^3 p}{(2 \pi)^3} \langle \tr_q 
\bigl( \Lambda_\alpha \gamma_5 S ( \omega_n + \epsilon_m, \mathbf{p} +  \mathbf{q}, A_4)\Lambda_\alpha \gamma_5 S (\epsilon_m, \mathbf{p}, A_4) \bigr) \rangle 
%\Pi_M^1(A_4)+\Pi_M^2(\omega_n,q, A_4). \label{Pi}
\nonumber \\
\label{Pipi}
\end{eqnarray}
for pseudo scalar fields for $\alpha = \pi, K, \eta, \eta'$ with the projection 
$\Lambda_\alpha$ onto the flavor channel $\alpha$.
Each terms on the right hand side of Eq.(\ref{p_mesons}) are multiplied by the corresponding degeneracy factor: 
3 for pions, 4 for kaons, {\it etc.} 
%contributions from pseudo scalar mesons; $\pi $, $K$, $\eta $ and $\eta '$. 
%The last four terms are from scalar mesons; $\sigma $, $\kappa $, $a_0$ and $f_0$. 
%The pre-factor of each $\mbox{ln}\mathcal{M}$ represents the number of degeneracy. 
%For example, since we have 3 pions, $\pi ^+$, $\pi ^-$ and $\pi^0$, the pre-factor of $\mbox{ln}\mathcal{M}_{\pi}$ is 3.
%Each $\mathcal{M}$ is given by Eqs.(\ref{M}) and (\ref{Pi}). 
%The index $M$ represents a kind of mesons. 

\begin{figure}[htbp]
\begin{center}
\includegraphics[clip,width=80mm]{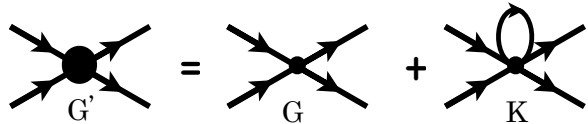}
\end{center}
\caption{
The effective four point coupling G' is a sum of the original coupling G and the one induced by the six point coupling in the presence of the condensate. 
}
\label{fig:eff_K}
\end{figure}

$G'_\alpha$ is the effective 4 point coupling, combination of the original four point coupling $G$ and the extra four point coupling generated from the six point coupling $K$ with appropriate weight for each channel, as indicated pictorially in Fig. \ref{fig:eff_K}. 
The explicit form of $G'$ will be given for each meson channel in the following subsections.
We note here that the effective four point coupling depends on the condensate $\langle \bar{q}_i q_j \rangle$ so that it depends on the temperature. 

%

%\begin{eqnarray}
%\Pi_M(\omega_n, q)=\Pi_M^1(A_4)+\Pi_M^2(\omega_n,q, A_4). \label{Pi}
%\end{eqnarray}

The discrete Matsubara frequency sum in (\ref{mpressure}) for  the mesonic correlation pressure 
%and the quark polarization (\ref{PiS, Pipi}) 
can be transformed by the method of contour integration to an integral along the positive real $\omega$ axis: 
\begin{eqnarray}
p_{M} & = & - \int \frac{d^3q}{(2\pi )^3} \int_0^\infty \frac{ d \omega}{2\pi i}  \left[ 1 + \frac{2}{e^{\beta \omega} -1} \right] 
 \Bigl\{ 3\mbox{ln}
\left[ \frac{ \mathcal{\tilde M}_{\pi }(\omega - i\delta, q)}{ \mathcal{\tilde M}_{\pi }(\omega + i\delta, q)} \right] 
\nonumber \\
& & 
+ 4 \mbox{ln} \left[ \frac{ \mathcal{\tilde M}_K (\omega - i\delta, q)}{ \mathcal{\tilde M}_K (\omega + i\delta, q)} \right]
+ \mbox{ln} \left[ \frac{ \mathcal{\tilde M}_\eta (\omega - i\delta, q)}{ \mathcal{\tilde M}_\eta (\omega + i\delta, q)} \right] 
+\mbox{ln} \left[ \frac{ \mathcal{\tilde M}_{\eta'} (\omega - i\delta, q)}{ \mathcal{\tilde M}_{\eta'} (\omega + i\delta, q)} \right]
 \nonumber \\
& & \qquad \qquad 
+ \mbox{ln} \left[ \frac{ \mathcal{\tilde M}_\sigma (\omega - i\delta, q)}{ \mathcal{\tilde M}_\sigma (\omega + i\delta, q)} \right]
+ 4 \mbox{ln} \left[ \frac{ \mathcal{\tilde M}_\kappa (\omega - i\delta, q)}{ \mathcal{\tilde M}_\kappa (\omega + i\delta, q)} \right]
\nonumber \\
& & \qquad \qquad \qquad + 3\mbox{ln} \left[ \frac{ \mathcal{\tilde M}_{a_0} (\omega - i\delta, q)}{ \mathcal{\tilde M}_{a_0} (\omega + i\delta, q)} \right] 
+\mbox{ln} \left[ \frac{ \mathcal{\tilde M}_{f_0} (\omega - i\delta, q)}{ \mathcal{\tilde M}_{f_0} (\omega + i\delta, q)} \right]
\Bigr\} \label{p_mesons} 
\label{mpressure1}
\end{eqnarray}
where  
\begin{equation}
\mathcal{\tilde{M}}_\alpha (\omega, q ) \equiv \mathcal{M}_\alpha (- i \omega, q ) =  \frac{1}{2K'_\alpha} - \tilde{\Pi}_\alpha ( \omega, q) .
\end{equation}
with 
\begin{eqnarray}
\tilde{\Pi}_\alpha (\omega, q) & = & \int \frac{d^3 p}{(2 \pi)^3} \int_0^\infty \frac{d \epsilon}{2 \pi i} \frac{-1}{e^{\beta \omega} + 1} 
\nonumber \\ 
& & \qquad \times \langle \tr_q 
\bigl( \Lambda_\alpha \gamma_5 S ( \epsilon + \omega, \mathbf{p} +  \mathbf{q}, A_4)\Lambda_\alpha \gamma_5 
S (\epsilon, \mathbf{p}, A_4) \bigr) \rangle 
%\Pi_M^1(A_4)+\Pi_M^2(\omega_n,q, A_4). \label{Pi}
\nonumber \\
\end{eqnarray}
where $S (\omega, \mathbf{p} )$ is the standard Feynman propagator for quark. 
We note that each logarithm in the integral (\ref{mpressure1}) is just the argument of $\tilde{\mathcal{M}}_\alpha (\omega + i\delta, q)$ multiplied by 2.  
%contains two types of diagrams drawn in Fig.\ref{fig:2point}.  The solid line represent quarks and the wave line do mesons.
%
In Eq. (\ref{mpressure1}), the first term in the bracket diverges in $\omega $ integral so that we need to introduce second cut-off parameter $\Lambda_b$ to suppress this divergence. In this work, we choose $\Lambda_b=\Lambda/2$ \cite{Nikolov:1996jj}.

\begin{figure}[htbp]
\begin{center}
\includegraphics[clip,width=30mm]{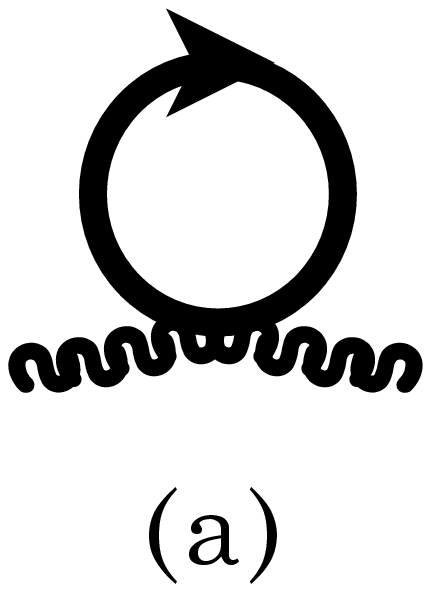}
\includegraphics[clip,width=50mm]{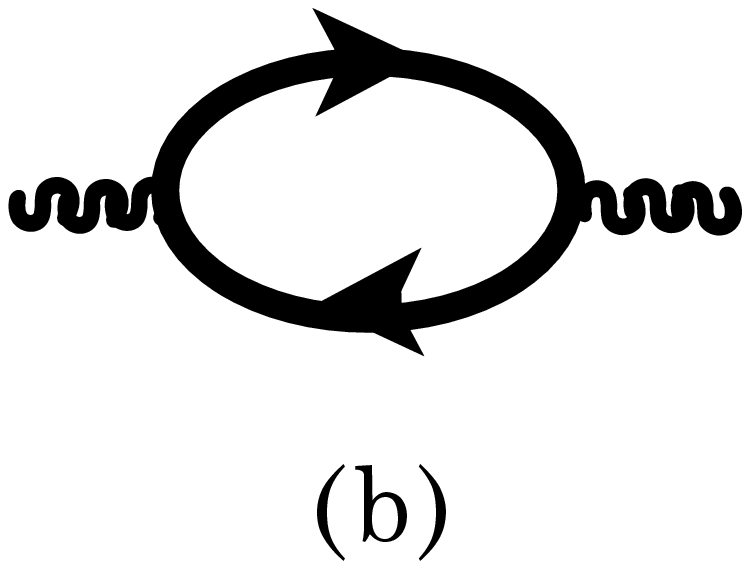}
\end{center}
\caption{
Meson self-energy, $\Pi _\alpha$, contains two terms, a non-dispersive contact term (a) and a dispersive term (b).
}
\label{fig:2point}
\end{figure}

For the computation of the correlation energy or pressure, it is convenient to decompose the function 
${\cal M} (\omega \pm i \delta , q ) $ into real part ${\cal M}_1( \omega, q) $ and imaginary part 
$ {\cal M}_2 ( \omega, q ) $: 
\begin{eqnarray}
{\cal M} ( \omega \pm i \delta , q ) & = & {\cal M}_1( \omega, q) \pm i {\cal M}_2 (\omega, q ) .
%{\cal F} ( \omega \pm i \delta , q ) =| {\cal F} (q, \omega \pm i \delta  ) | e^{\pm i \phi (q, \omega) } 
\end{eqnarray}
The imaginary part $ {\cal M}_2 ( \omega, q ) $ becomes non zero in the kinematical region of $(\omega, q)$ which allows a pair 
excitation of quark and antiquark  or scattering of a thermally excited quark into an unoccupied state, as signified by non-zero imaginary part of the function $\mathcal{F}^{\rm pair} (\omega \pm i \delta, q )$ and  $\mathcal{F}^{\rm scat} (\omega \pm i \delta, q )$, respectively, given explicitly in Appendix B.   
The long-lived meson collective mode exists when the real part ${\cal M}_1( \omega, q) $ vanishes in the region where the imaginary part also vanishes.

\subsection{Pseudo scalar mesons: $\pi $, $K$, $\eta $, $\eta '$}

In this section, we consider the contribution to pressure from pseudo scalar mesons which make a nonet in the SU(3) flavor space. From Eq.(\ref{p_mesons}), we see the pressure of them is written as the sum of contributions from each meson.
With SU(3) flavor symmetry breaking, keeping the isospin SU(2) symmetry intact, pseudo scalar mesons are classified in four kinds by the difference of their masses. 

The difference of four mesons appears in the meson self-energy due to quark polarization $\Pi_\alpha$ and the effective four point coupling $G'_\alpha$. 
$\Pi_\alpha$ of pseudo scalar mesons are written in general by 
\begin{eqnarray}
\Pi _{PS}=\Pi _{PS}^1(A_4)+\Pi_{PS}^2(\omega_n , q, A_4)
\end{eqnarray}
where
\begin{eqnarray}
\Pi _{PS}^1(A_4) & = &-2T \sum_n\mbox{tr}_c\int \frac{d^3p}{(2\pi )^3}\frac{1}{2}
\Bigl( \frac{1}{(\epsilon _n + gA_4)^2+E_i(p)^2}+\frac{1}{(\epsilon _n+gA_4)^2+E_j(p)^2}\Bigr) 
\nonumber \\
& = &  - \bigl( \frac{\langle \langle \bar{q}_i q_i \rangle \rangle}{M_i} + \frac{\langle \langle \bar{q}_j q_j \rangle \rangle}{M_j} \bigr)
\end{eqnarray}
is a non-dispersive contact component of the meson self-energy as shown diagramaticaly in Fig. 7 (a), while 
\begin{eqnarray}
\Pi _{PS}^2(\omega_n, q, A_4) = \bigl( \omega_n^2+q^2+(M_i-M_j)^2\bigr) F_{ij} (\omega_n, q, A_4) .
\end{eqnarray}
with
\begin{eqnarray}
F_{ij} (\omega_n, q , A_4) & = & 2 T \sum_{n'} \mbox{tr}_c\int \frac{d^3p}{(2\pi )^3}
\frac{1}{[(\epsilon_{n'} + gA_4)^2+E_i (p)^2]}
\nonumber \\
& & \qquad \qquad \qquad \times
\frac{1}{ [(\epsilon_{n'} + gA_4 + \omega_n)^2+E_j (p+q)^2]} \ \ \ \ \ 
\end{eqnarray}
is a dispersive component shown in Fig. 7 (b). 
In these expression, $i, j (= u, d, s)$ indicate the flavors of quark and antiquarks constituting the pseudo scalar meson. 

The dispersive part of the meson self-energy needs to be analytically continued to a Fourier transform of the real time expression in order to find the dispersion relation for collective meson modes. 
The detail of computation is given in the Appendix A of our previous paper\cite{Yamazaki:2012ux}. 
Here we present a result of such computations for pseudo scalar mesons:
\begin{eqnarray}
\tilde{\Pi}^2_{PS} (\omega, q, A_4) = \bigl( - \omega^2+q^2+(M_i-M_j)^2\bigr) \mathcal{F}_{ij} (\omega, q, A_4) .
\end{eqnarray}
where 
\begin{eqnarray}
\mathcal{F}_{ij} (\omega, q , A_4) & = & \mathcal{F}_{ij}^{\bf scat} (\omega, q , A_4) +
 \mathcal{F}_{ij}^{\bf pair} (\omega, q , A_4) 
\end{eqnarray}
with the scattering term 
\begin{eqnarray}
\mathcal{F}_{ij}^{\bf scat} (\omega, q , A_4) & = & 
\int \frac{d^3p}{(2\pi )^3} \frac{1}{2E_i (p) 2E_j (p+q)}  \left( \frac{1}{\omega + E_i (p) - E_j (p+q)} \right.
\nonumber \\
& & \qquad  \qquad 
\left. - \frac{1}{\omega - E_i (p) + E_j (p+q)} \right)
\nonumber \\
& & \qquad 
\times \tr_c \left[ f ( E_i (p) - ig A_4) - f (E_j (p+q) - i g A_4) \right]
\end{eqnarray}
and the pair creation and annihilation term 
\begin{eqnarray}
\mathcal{F}_{ij}^{\bf pair} (\omega, q , A_4) & = & 
\int \frac{d^3p}{(2\pi )^3} \frac{1}{2E_i (p) 2E_j (p+q)}  \left( \frac{1}{\omega + E_i (p) + E_j (p+q)} \right.
\nonumber \\
& & \qquad  \qquad 
\left. - \frac{1}{\omega - E_i (p) - E_j (p+q)} \right)
\nonumber \\
& & \qquad 
\times \tr_c \left[ 1 - f ( E_i (p) - ig A_4) - f (E_j (p+q) - i g A_4) \right]
\nonumber \\
\end{eqnarray}
These functions possess singularities when $\omega = \pm ( E_i (p) - E_j (p+q) )$ is fulfilled for the scattering term and
$\omega =  \pm (E_i (p) + E_j (p+q)) $ for the pair term corresponding to the real excitations of the medium. 
Note that the effect of the background gauge field $A_4$ cancels for these excitation which are totally color singlet. 
However, the gauge field still appears in the distribution functions of the quark quasiparticles as phase factor in exactly the same way in the mean field calculation.  
We replace these phase factor by Polyakov loops and then by the statistical average, e. g. 
\begin{equation}
\tr_c \left[ f ( E_i (p) - ig A_4) - f (E_j (p+q) - i g A_4) \right] \to \left[ f_\Phi ( E_i (p) ) - f _\Phi (E_j (p+q) ) \right]
\end{equation}
for the statistical averages of each component,
\begin{eqnarray}
\mathcal{F}_{ij}^{\bf scat} (\omega, q) & = & \langle \mathcal{F}_{ij}^{\bf scat} (\omega, q , A_4) \rangle , \quad
\mathcal{F}_{ij}^{\bf pair} (\omega, q)  =  \langle \mathcal{F}_{ij}^{\bf pair} (\omega, q , A_4) \rangle .
\end{eqnarray}

Having discussed the generic results for pseudo scalar mesons, we now present the explicit form for each pseudo scalar mesons starting from pions. 
With the SU(2) isospin symmetry, $u$-quark and $d$-quark are degenerate so that contribution to pressure from three pions $\pi ^{+}$, $\pi ^{-}$ and $\pi ^{0}$ are identical.  
\begin{eqnarray}
\mathcal{\tilde{M}}_{\pi } ( \omega , q ) = \frac{1}{2 G'_\pi} - \tilde{\Pi} _{\pi }^1  - \tilde{\Pi} _{\pi}^2 (\omega, q) 
\label{Mpion}
\end{eqnarray}
where 
\begin{eqnarray}
G'_\pi \equiv G'_1 = G'_2 = G'_3 = G - \frac{K}{2} \langle \langle \bar{s}s \rangle \rangle , 
\label{K3} 
\end{eqnarray}
is  the effective four point coupling for pions and non-dispersive and dispersive parts of the meson self-energy for pions are given by 
\begin{eqnarray}
%\tilde{\Pi} _{\pi } (\omega, q) 
& &\tilde{\Pi} _{\pi }^1  =   \langle \Pi _{\pi}^1 ( A_4) \rangle
= - 2 \frac{\langle \langle \bar{u} u \rangle \rangle}{M_u}
%\tilde{\Pi} _{\pi } (\omega, q) 
\nonumber \\
& & \tilde{\Pi} _{\pi}^2 (\omega, q)  =   \langle \tilde{\Pi} _{\pi}^2(\omega, q, A_4) \rangle 
= (-\omega^2+q^2) \mathcal{F}_{\pi }(\omega, q)
%\nonumber \\
\label{Pipi}
\end{eqnarray}
An explicit form of $ \mathcal{F}_{\pi }(\omega, q)$ is given in Appendix A. 
%\begin{eqnarray} 
%\mathcal{F}_{\pi }(\omega, q ) = 
%2T\sum_{n'}\mbox{tr}_c\int \frac{d^3p}{(2\pi )^3}
%\frac{1}{[(\epsilon_{n'}+gA_4)^2+E_u(p)^2] [(\epsilon_{n'}+gA_4+\omega_n)^2+E_u(p+q)^2]} \ \ \ \ \ 
%\end{eqnarray}

Comparing (\ref{Pipi}) with the gap equation for $M_u$, the first two terms on the right hand side of (\ref{Mpion}) can be transformed into a simpler form: 
%second term,  $\Pi _{\pi }^1$,  is eliminated by gap equations and then we get $\mathcal{M}_\pi $ as a function of  meson energy $\omega $ and momentum $q$.
%\begin{eqnarray}
%\frac{\langle \bar{u}u\rangle }{M_u}=-\frac{M_u-m_u}{4K_3M_u}. \label{u_loop}
%\end{eqnarray}
%
\begin{eqnarray}
\tilde{\mathcal{M}}_{\pi } (\omega, q) = (-\omega^2 + q^2) \mathcal{F}_{\pi }(\omega, q ) + \frac{m_u}{2G'_\pi M_u} 
\label{p_pi}
\end{eqnarray}
We note that in the limit, $K=0$,  $\tilde{\mathcal{M}}_{\pi } $ coincides with our previous results with the two flavor model\cite{Yamazaki:2012ux}.

Similarly, for  kaons, we find, 
\begin{eqnarray}
\tilde{\mathcal{M}}_K (\omega, q ) = \frac{1}{2G'_K} - \langle \tilde{\Pi} _K^1 \rangle - \langle \tilde{\Pi} _K^2 (\omega, q) \rangle
\label{MK}
\end{eqnarray}
where
\begin{eqnarray}
G'_K \equiv = G'_4 = G'_5 =G'_6 = G'_7 =  G - \frac{K}{2} \langle \langle \bar{u} u \rangle\rangle , 
\label{K6} 
\end{eqnarray}
and 
\begin{eqnarray}
&& \langle \tilde{\Pi}_{K}^1 \rangle  =  - \bigl( \frac{\langle \langle \bar{u} u \rangle \rangle}{M_u} +
\frac{\langle \langle \bar{s} s \rangle \rangle}{M_s} \bigr) \\
%  2 T\sum _n\mbox{tr}_c\int \frac{d^3p}{(2\pi )^3}\frac{1}{2}
%\Bigl( \frac{1}{(\epsilon_n+gA_4)^2+E_u(p)^2}+\frac{1}{(\epsilon_n+gA_4)^2+E_s(p)^2}\Bigr) \ \ \ \ \ 
%\end{eqnarray}
%\begin{eqnarray}
& & \langle \tilde{\Pi}_{K}^2 (\omega, q) \rangle =  (- \omega^2 + q^2 + (M_s-M_u)^2) \mathcal{F}_K (\omega, q)
\end{eqnarray}
with $\mathcal{F}_K (\omega, q)$ given in the Appendix A.
%\begin{eqnarray}
%\mathcal{F}_{K}(\omega_n, q , A_4)=2T\sum_{n'}\mbox{tr}_c\int \frac{d^3p}{(2\pi )^3}
%\frac{1}{[(\epsilon_{n'}+gA_4)^2+E_u(p)^2] [(\epsilon_{n'}+gA_4+\omega_n)^2+E_s(p+q)^2]} \ \ \ \ \ 
%\end{eqnarray}
By making use of the gap equation for $M_s$, (\ref{MK}) is transformed to
%After performing the same method as the case of pions, $\Pi _K^1$ can be eliminated by the gap equations. 
%\begin{eqnarray}
%\frac{\langle \bar{s}s\rangle }{M_s}=\frac{M_u-m_u}{4K_6M_s}-\frac{G}{K_6M_s}\bigl( \langle \bar{u}u\rangle -\langle \bar{s}s\rangle \bigr) \\
%\frac{\langle \bar{u}u\rangle }{M_u}=\frac{M_s-m_s}{4K_6M_u}-\frac{G}{K_6M_u}\bigl( \langle \bar{s}s\rangle -\langle \bar{u}u\rangle \bigr) 
%\end{eqnarray}
%As a result, we get $\mathcal{M}_K$: 
\begin{eqnarray}
\tilde{\mathcal{M}}_K ( \omega, q) & = & (- \omega^2 + q^2 + (M_u-M_s)^2) \mathcal{F}_K (\omega, q) \nonumber \\
& &\qquad +\frac{1}{2G'_K}
-\frac{M_u-m_u}{4G'_K M_s}-\frac{M_s-m_s}{4G'_K M_u} 
\nonumber \\
& & \qquad \qquad 
- \frac{G}{G'_K}\Bigl( \frac{\langle \langle \bar{u}u \rangle \rangle -\langle \langle \bar{s}s \rangle \rangle}{M_s}
+\frac{\langle \langle \bar{s}s\rangle \rangle -\langle \langle \bar{u}u \rangle \rangle }{M_u}\Bigr)  %\nonumber 
\end{eqnarray}
Although it looks rather complicated, this result coincides with that for pions when the flavor $SU(3)$ symmetry becomes  exact.

Next we consider $\eta$ and $\eta'$ mesons. 
If there were no Kobayashi-Masukawa-'t Hooft interaction, in other words if $U(1)_A$ symmetry is not broken,  masses of these two mesons are same. 
Both $\eta$ and $\eta'$ mesons are mixtures of a flavor singlet $\eta^0$ and  a flavor octet $\eta^8$.  
Without mixing, we find for $\eta^8$, 
\begin{eqnarray}
\tilde{\mathcal{M}}_{\eta^8} (\omega, q ) & = & \frac{1}{2G'_{\eta^8}} - \langle \tilde{\Pi} _{8}  (\omega, q )  \rangle 
\nonumber \\
& = & (- \omega^2 + q^2 ) \mathcal{F}_{\eta^8} (\omega, q) \nonumber \\
%\nonumber \\
& & \qquad 
+ \frac{1}{2G'_{\eta^8}}-\frac{2}{3}\Bigl[ 
\frac{1}{3G'_8}\Bigl( \frac{M_s-m_s}{M_u}+2\frac{M_u-m_u}{M_s}\Bigr) \Bigl( \frac{\langle \langle \bar{s}s\rangle \rangle}{4\langle \langle \bar{u}u\rangle \rangle} -1\Bigr) 
\nonumber \\
& & \qquad + \frac{G}{3G'_8}\Bigl\{ \frac{1}{M_u}\Bigl( \frac{\langle \langle \bar{s}s\rangle \rangle\langle \langle \bar{s}s\rangle \rangle}{\langle \langle \bar{u}u\rangle \rangle}-4\langle \langle \bar{s}s\rangle \rangle+3\langle \langle \bar{u}u\rangle \rangle\Bigr) 
%\nonumber \\
%& & \qquad \qquad 
+\frac{8}{M_s}\bigl( \langle \langle \bar{s}s\rangle \rangle -\langle \langle \bar{u}u\rangle \rangle \bigr) \Bigr\}  \Bigr] 
\nonumber \\
\end{eqnarray}
where 
\begin{eqnarray}
G'_{\eta^8}=G+\frac{K}{6}\bigl( \langle \langle \bar{s}s\rangle \rangle -4\langle \langle \bar{u}u\rangle \rangle \bigr) \label{K8}
\end{eqnarray}
and for $\eta^0$,
\begin{eqnarray}
\tilde{\mathcal{M}}_{\eta^0} & = & \frac{1}{2G'_{\eta^0}}- \langle \tilde{\Pi} _{00} (\omega, q) \rangle
\nonumber \\
& = & (- \omega^2 + q^2 ) \mathcal{F}_{\eta^0} (\omega, q) \nonumber \\
& & \qquad 
-\frac{2}{3}\Bigl[  \frac{1}{3G'_{\eta^0}}\Bigl( 2\frac{M_s-m_s}{M_u}+\frac{M_u-m_u}{M_s}\Bigr) \Bigl( \frac{\langle \langle \bar{s}s\rangle \rangle}{2\langle \langle \bar{u}u\rangle \rangle} +1\Bigr) 
\nonumber \\
& & \qquad  + 
\frac{G}{3G'_{\eta^0}}\Bigl\{ \frac{1}{M_u}\Bigl(4 \frac{\langle \langle \bar{s}s\rangle \rangle \langle \langle \bar{s}s\rangle \rangle}{\langle \langle \bar{u}u\rangle \rangle}+8\langle \langle \bar{s}s\rangle \rangle+6\langle \langle \bar{u}u\rangle \rangle\Bigr) \Bigr] 
%\nonumber \\
%& & \qquad \qquad \qquad 
+\frac{1}{M_s}\bigl( 5\langle \langle \bar{s}s\rangle \rangle +4\langle \langle \bar{u}u\rangle \rangle \bigr) \Bigr\}  
\nonumber \\
\end{eqnarray}
where
\begin{eqnarray}
G'_{\eta^0}=G+\frac{K}{3}\bigl( \langle \langle \bar{s}s\rangle \rangle +2\langle \langle \bar{u}u\rangle \rangle \bigr) \label{K0}
\end{eqnarray}

To check these results for $\eta^8$ and $\eta^0$, we inspect these formulae at extreme cases. 
First, we consider the case with exact $SU_L (3) \times SU_R (3)$ symmetry, taking $m_u = m_d = m_s = 0$, but with finite Kobatashi-Masukawa-'t Hooft coupling K. 
In this case, $\eta^8$ becomes a massless Nambu-Goldstone mode, while $\eta^0$ becomes massive mode. 
If further set we $K=0$  keeping the $SU(3)$ chiral symmetry,  $\eta^0$ also becomes a massless NG mode.  
Regarding the mixing of $\eta^0$ and $\eta^8$, we need an effective coupling $G'_{08}$;
\begin{eqnarray}
G'_{08}=G+\frac{\sqrt{2}K}{6}(\langle \langle \bar{u}u\rangle \rangle -\langle \langle \bar{s}s\rangle \rangle )
\end{eqnarray}
and $\langle  \tilde{\Pi}_{08}(\omega , q)\rangle $. 

\begin{figure}[htbp]
\begin{center}
\includegraphics[clip,width=80mm, angle=270]{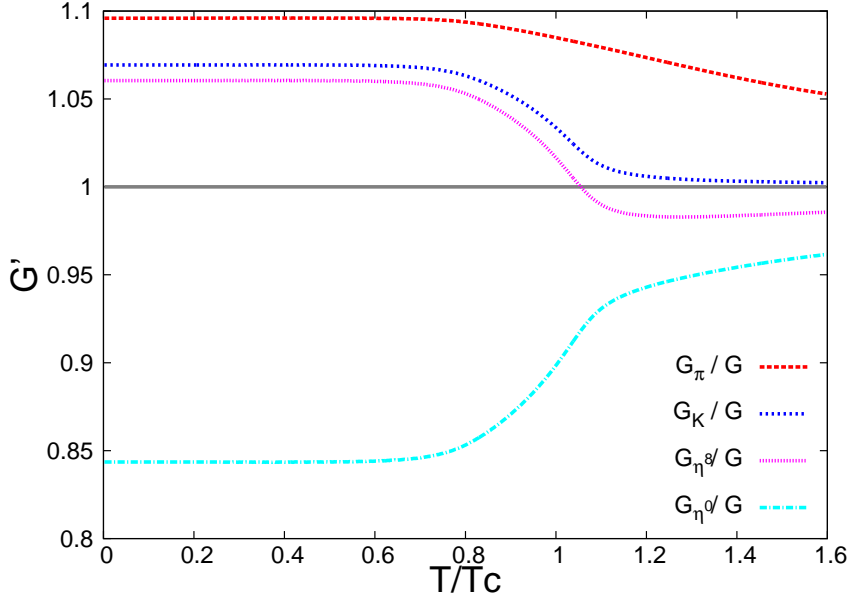}
\end{center}
\caption{
Temperature dependence of the effective four point coupling $G'$ for pions (the dotted red line), kaons (the dotted blue line),  
 $\eta^8$ (the dotted pink line), and $\eta^0$ (the dashed light blue line).
% thescaled by the original four point coupling $G$ as a function of temperature. Red line: $G_{\pi}/G$.  Green line: $G_K/G$.  Blue line: $G_{\eta^8}/G$. Pink line: $G_{\eta^0}/G$.
The temperature is scaled by the pseudo critical temperature $T_c$ computed by the mean field approximation. 
}
\label{fig:G}
\end{figure}

We show in Fig.\ref{fig:G} the effective coupling $G'$ scaled by four point coupling $G$ as a function of $T/T_c$ in order to check the effect of the Kobayashi-Masukawa-'t Hooft interaction. Each $G'$ indirectly depends on temperature through chiral condensates. 
Although this interaction is introduced to make the mass splitting of $\eta^8$ and $\eta^0$, it also influences on the couplings of $\pi$ and $K$. One can see the effective coupling of $\pi $ doesn't change so much compared with others as temperature increases because of  the moderate change of $\langle \bar{s}s\rangle $.  
Another feature point is that only $G'_{\eta^0}$ becomes smaller than the original four point coupling $G$.   Others become larger than $G$ because of the negative effect of the second terms in Eqs. (\ref{K3}),(\ref{K6}) and (\ref{K8}).

\subsection{Scalar mesons}

We found for scalar mesons, 

\begin{eqnarray}
\tilde{\mathcal{M}}_S (\omega, q ) = \frac{1}{2G'_S} - \langle \tilde{\Pi} _S^1 \rangle - \langle \tilde{\Pi} _S^2 (\omega, q) \rangle
\label{MK}
\end{eqnarray}
where
\begin{eqnarray}
 G'_S = 
 \begin{cases}
& G + \frac{\displaystyle K}{\displaystyle 2} \langle \langle \bar{s} s \rangle \rangle  \qquad \qquad {\rm for} \quad a_0 \\
\\
& G + \frac{\displaystyle K}{\displaystyle 2} \langle \langle \bar{u} u \rangle \rangle  \qquad \qquad {\rm for} \quad \kappa \\
\\
& G - \frac{\displaystyle K}{\displaystyle 6} \left( \langle \langle \bar{s} s \rangle \rangle -4 \langle \langle \bar{u} u \rangle \rangle \right) \qquad \qquad {\rm for} \quad \sigma \\
\\
& G - \frac{\displaystyle K}{\displaystyle 3} \left( \langle \langle \bar{s} s \rangle \rangle + 2\langle \langle \bar{u} u \rangle \rangle \right) \qquad \qquad {\rm for} \quad f_0 
\end{cases}
\end{eqnarray}
\begin{eqnarray}
&& \langle \Pi_{S}^1 \rangle  =  - \bigl( \frac{\langle \langle \bar{u} u \rangle \rangle}{M_i} +
\frac{\langle \langle \bar{s} s \rangle \rangle}{M_j} \bigr) = \langle \Pi_{PS}^1 \rangle   \\
%  2 T\sum _n\mbox{tr}_c\int \frac{d^3p}{(2\pi )^3}\frac{1}{2}
%\Bigl( \frac{1}{(\epsilon_n+gA_4)^2+E_u(p)^2}+\frac{1}{(\epsilon_n+gA_4)^2+E_s(p)^2}\Bigr) \ \ \ \ \ 
%\end{eqnarray}
%\begin{eqnarray}
& & \langle \tilde{\Pi}_{S}^2 (\omega, q) \rangle =  (- \omega^2 + q^2 + (M_i + M_j)^2) \mathcal{F}_S (\omega, q)
\end{eqnarray}
with $\mathcal{F}_S (\omega, q) = \mathcal{F}_{PS} (\omega, q) $.

The only but important difference between $\Pi_{PS}^2$ for pseudo scalars and $\Pi_S^2$ for scalars is 
in the first term of the dispersive parts where the combination $M_i + M_j$ for scalar case appears instead of 
$M_i - M_j$.  
Hence if $M_i = M_j$, this factor generates a massless Nambu-Goldstone mode for pseudo scalar cases, 
while scalar mesons becomes massive with its mass given by $M_S = 2 M_i$.

\subsection{Numerical results: mesonic correlation pressure and melting of collective meson modes}

\begin{figure}[htbp]
\begin{center}
\includegraphics[clip,width=80mm, angle=270]{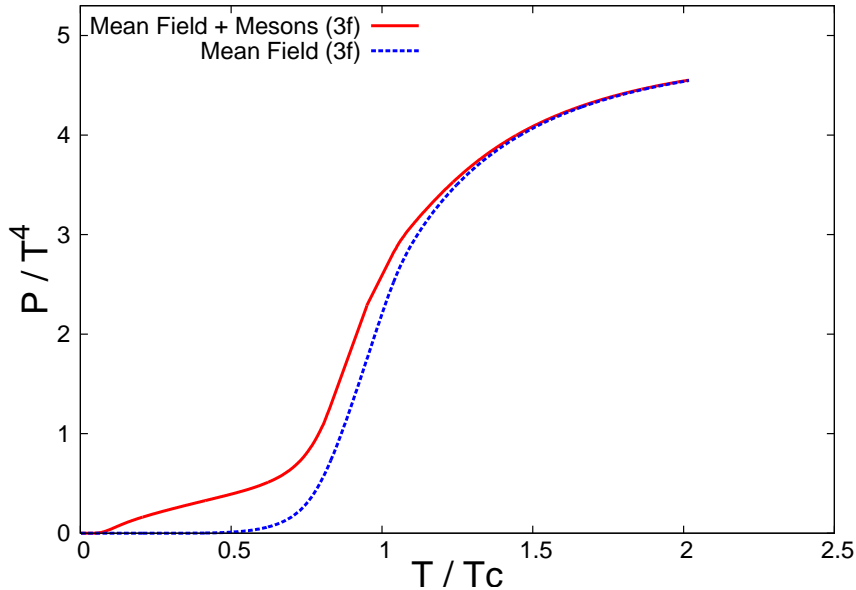}
\end{center}
\caption{
Pressure scaled by $T^4$ as a function of the temperature $T$ scaled by the pseudo critical temperature $T_c$. 
The dashed blue line is the result of the mean field approximation,  consisting of quark pressure and gluon pressure,
while the solid red line contains also the pressure from mesonic correlations in the pion, kaon, and $\sigma$ meson channel. 
% in the mean field approximation : sum of the contributions from the mean field and mesonic correlations.  Blue line: without mesonic correlations.  The pseudo critical temperature $T_c$ is computed by 
%the mean field approximation. 
}
\label{fig:EOS_3f}
\end{figure}

We present the result of the pressure in Fig.\ref{fig:EOS_3f} calculated by the previous method.  
We only include $\pi$, $K$ and $\sigma$ in this calculation since we expect that other more massive mesons 
would not contribute much at low temperatures to the pressure.
% compared with $\pi$, $K$ and $\sigma$. 

At low temperatures,  the pressure is dominated by mesonic correlations, especially pions and kaons. 
As temperature increases, it approaches to the quark mean field pressure.
% and corresponds to it at high temperatures. 
It means that mesonic collective modes melt as temperature increases and dissolve into quarks eventually. As compared with the two flavor case, contributions of kaons is added on the pressure of two flavor which are dominated by a pion gas at low temperatures and becomes continuously a gas of $u$, $d$ and $s$ quarks at high temperatures. 

\begin{figure}[htbp]
\begin{center}
%\resizebox*{!}{4.7cm}{
%\vspace*{0.5cm} 
\includegraphics[clip,width=50mm, angle=270]{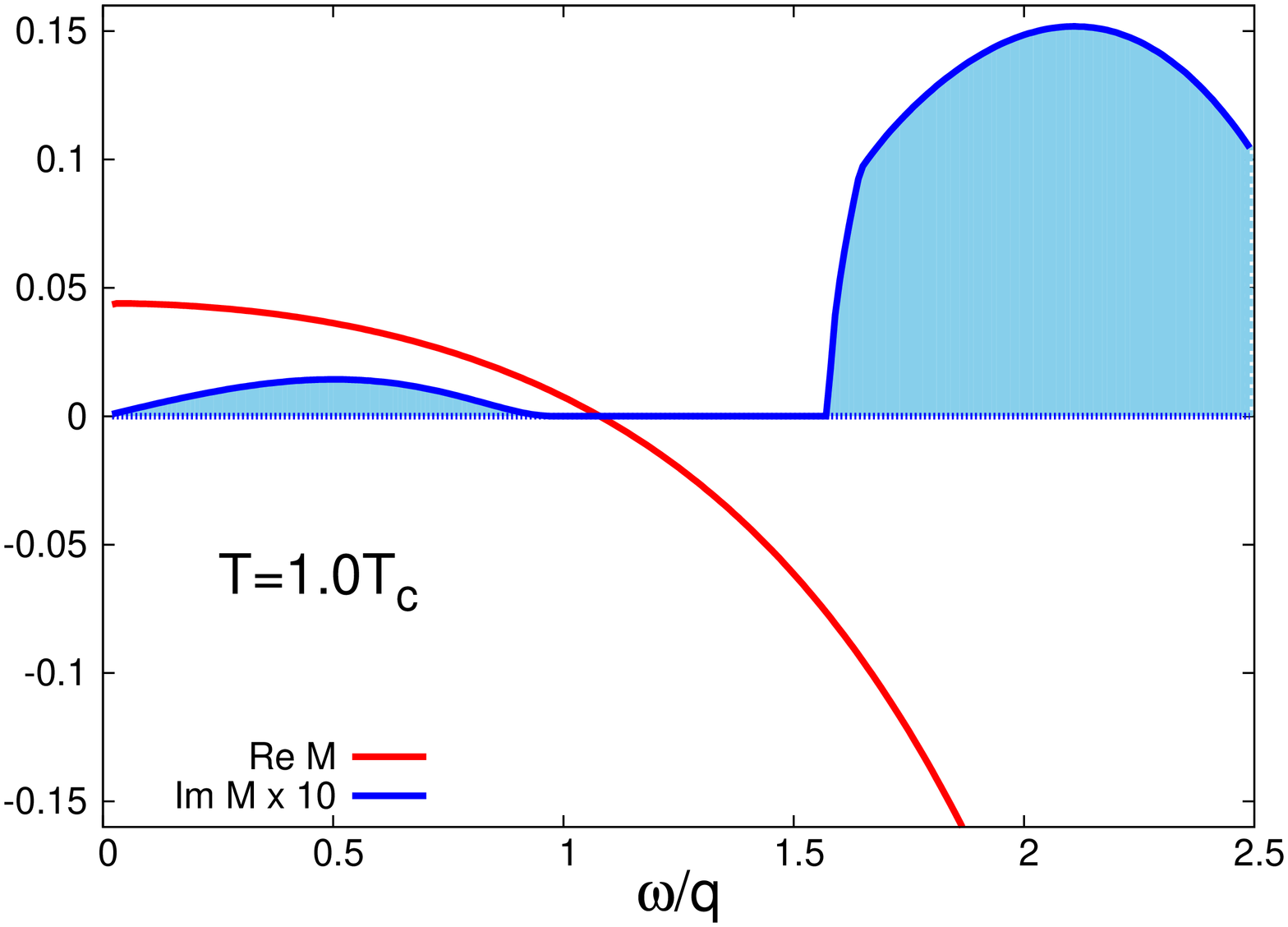}
%\vspace*{0.5cm} 
\includegraphics[clip,width=50mm, angle=270]{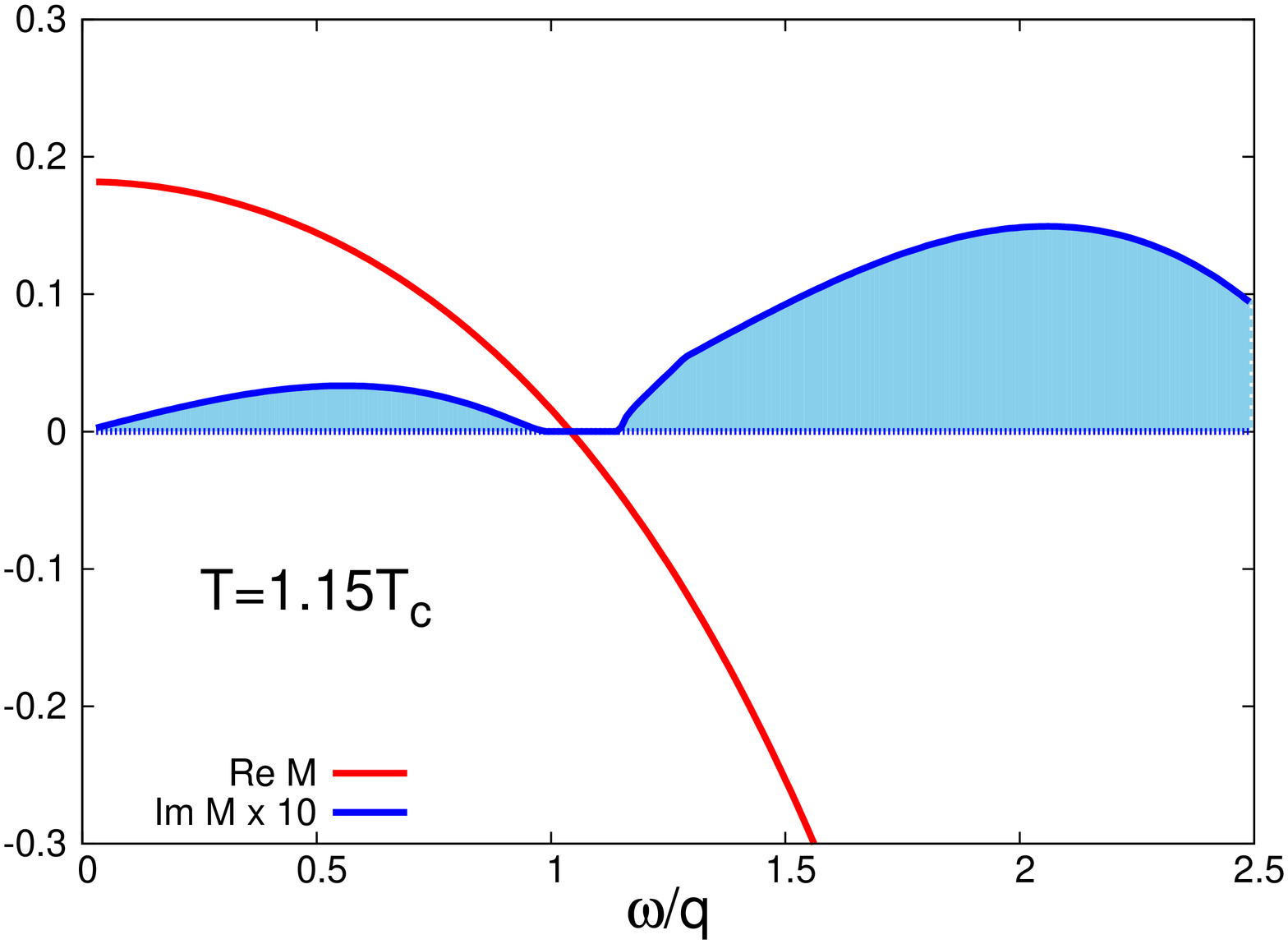}
%\vspace*{0.5cm} 
\includegraphics[clip,width=50mm, angle=270]{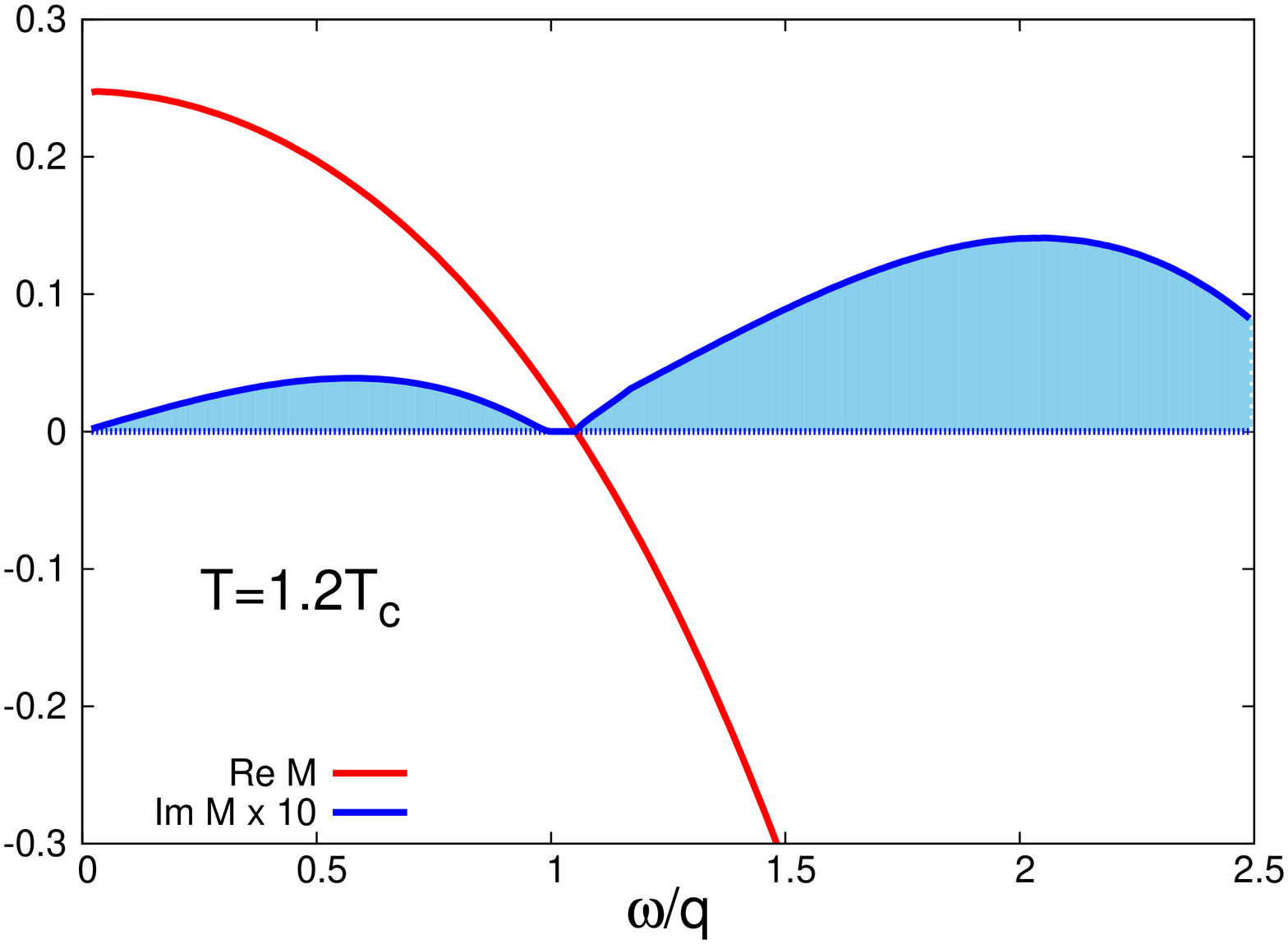}
\end{center}
\caption{
The real part $\mathcal{M}_1$ (the solid red line) and the imaginary part $\mathcal{M}_2$ (the solid blue line) of $\mathcal{M} (\omega + i \delta, q) $ for pion is plotted as a function of $\omega$ scaled by $q$ at several temperature near in the crossover region.  The non-zero imaginary part are indicated by the filled light blue areas in the space-like $\omega/q <1$ and in the time-like $\omega/q >1$ regions.  
When $\mathcal{M}_1$ vanishes in the region where $\mathcal{M}_2$ also vanishes, the long-lived pion collective mode exists, as 
seen below $T=1.15 T_c$ clearly.   Pion collective mode is absorbed into the continuum of the pair excitations at $T=1.2 T_c$.
}
\label{fig:coll_PI}
\end{figure}

In order to study until which temperature mesons as collective modes persist, 
we plot the real part  and the imaginary part of $\mathcal{M} (\omega, q)$ as a function of $\omega $ scaled by $q$ at several temperatures near $T_c$. The conditions of isolated meson poles are given by the vanishing both real part $\mathcal{M}_1$ and imaginary part $\mathcal{M}_2$: in particular the condition $\mathcal{M}_1(\omega, q)=0$ determines the dispersion relation of collective modes. 
$\mathcal{M}_2$ corresponds to the excitations of quarks and antiquarks as a continuum.   
Therefore collective meson modes exist when both the real part and the imaginary part of 
$\tilde{\mathcal{M}} (\omega + i \delta, q)$ vanish at the same time, the former condition determines the dispersion relation of the meson and the latter guarantees infinite lifetime. We write the real part and imaginary part of $\mathcal{M}(\omega, q)$ in Appendix B. 

\begin{figure}[htbp]
\begin{center}
%\resizebox*{!}{4.7cm}{
%\vspace*{1cm} 
\includegraphics[clip,width=50mm, angle=270]{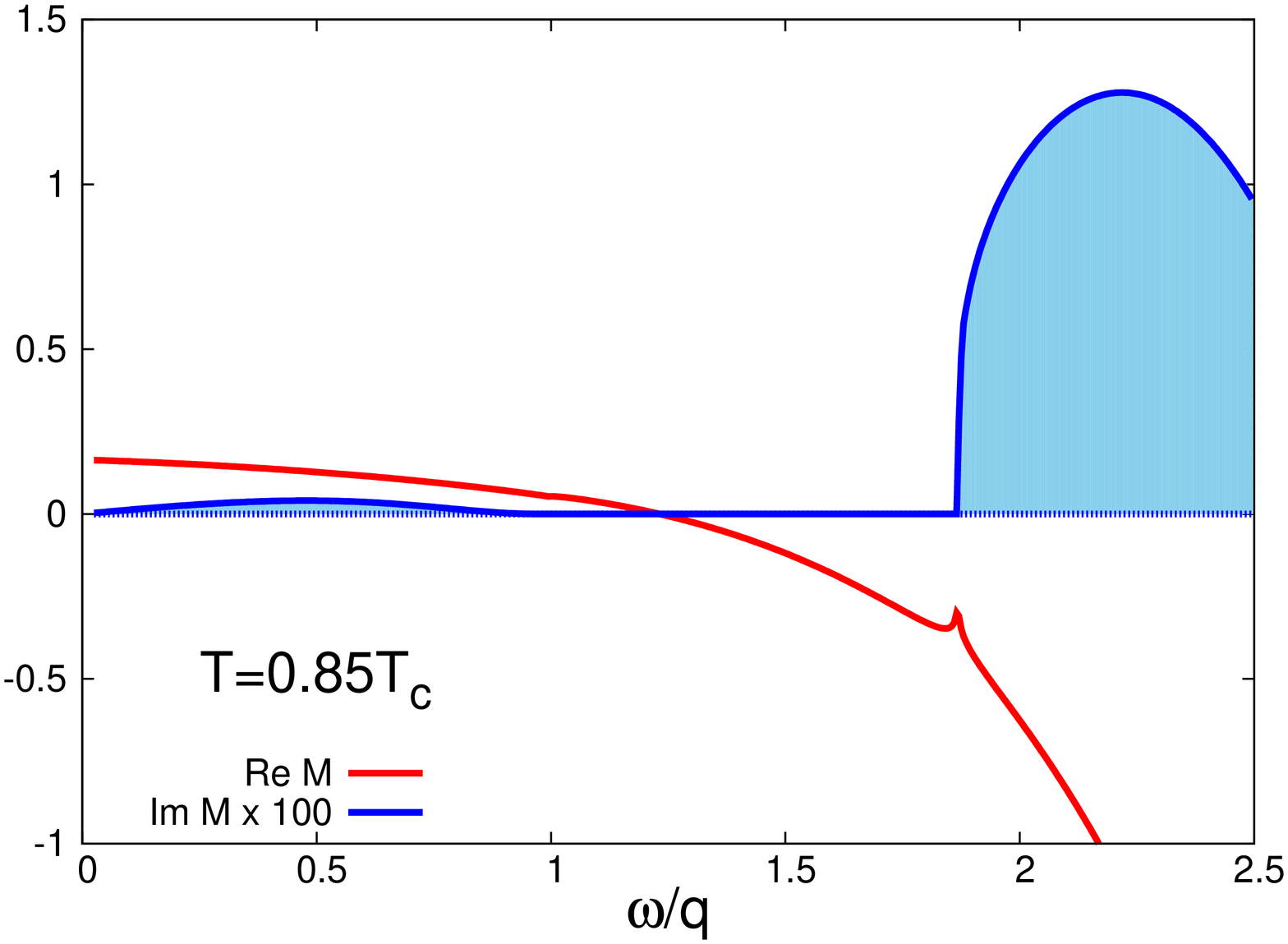}
%\vspace*{1cm} 
\includegraphics[clip,width=50mm, angle=270]{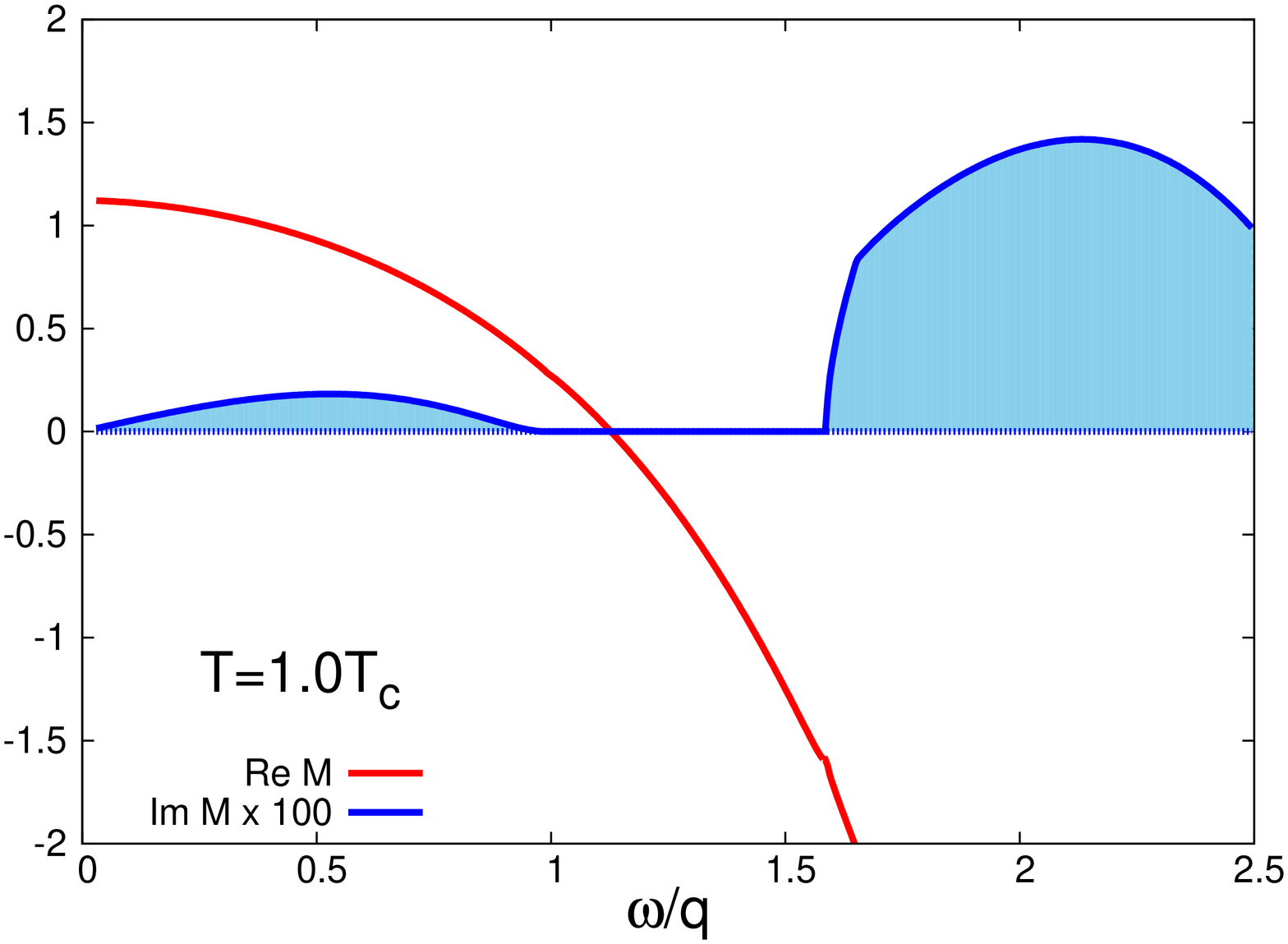}
%\vspace*{1cm} 
\includegraphics[clip,width=50mm, angle=270]{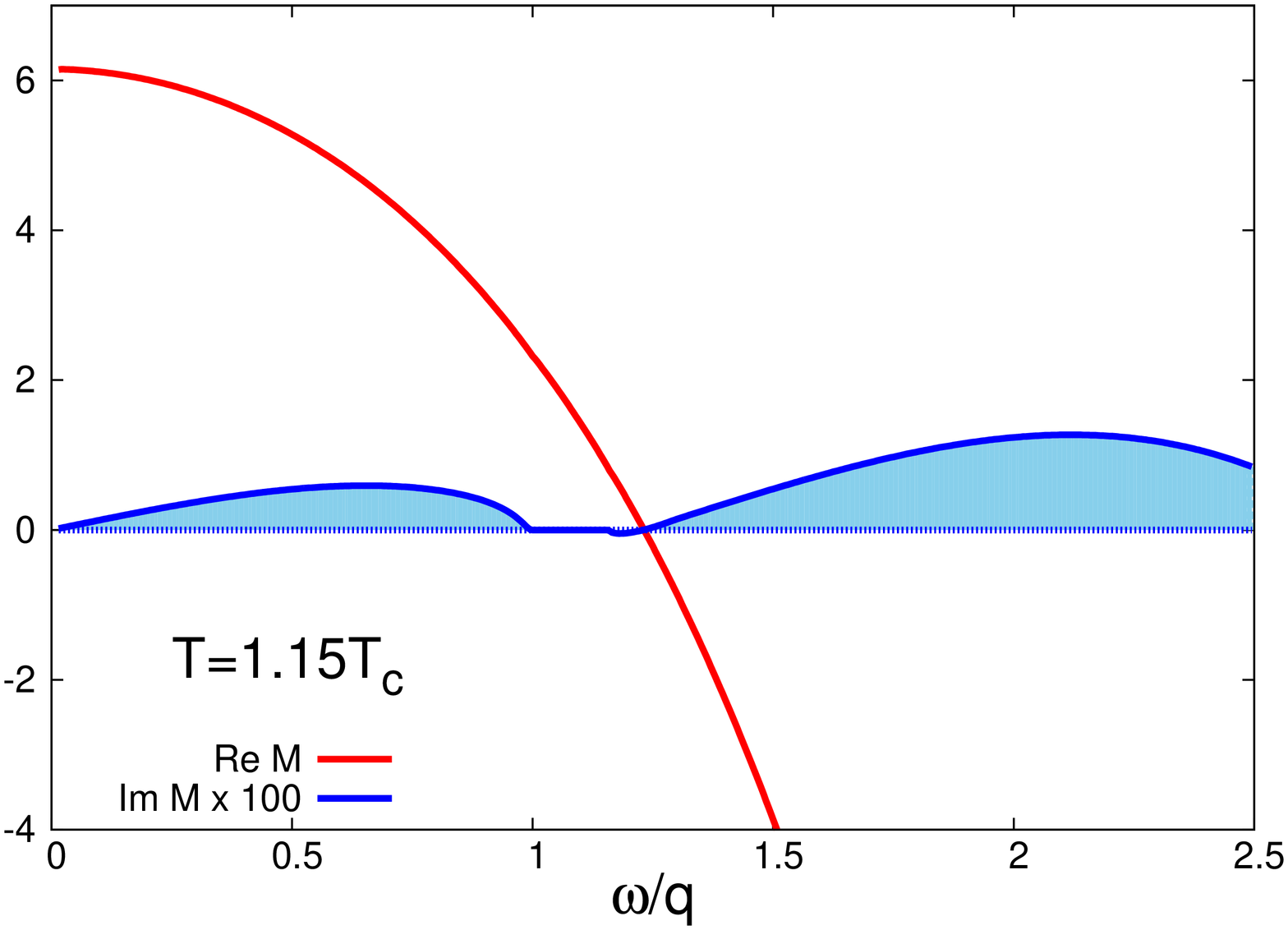}
\end{center}
\caption{
The same plots as the Fig. (\ref{fig:coll_PI}) for kaons. 
The collective kaon mode is absorbed into the pair continuum at $T=1.15 T_c$, slightly below the pion melting temperature.
%The real part of $\mathcal{M}$ and imaginary part of$ \mathcal{M}$ for kaon as a function of $\omega$ scaled by $q$. 
}
\label{fig:coll_K}
\end{figure}

We plot each component as a function $\omega/q$ in Fig.\ref{fig:coll_PI}  for pion and Fig. \ref{fig:coll_K}  for kaon at three different temperatures around $T_c$. 
These panels are showing whether collective modes exist or not for pion and for kaon. In the shadowed areas, where the imaginary part $\mathcal{M}_2$ has finite value, the continuum of quark and anti-quark exists. The region where the imaginary part is finite in time-like  goes down as temperature increases because of a decrease of constituent quark masses along with chiral symmetry recovering. 

In Fig. \ref{fig:coll_PI},  there is a point where both real and imaginary parts become zero until $T = 1.15 T_c$, but when T reaches $1.2T_c$ such a point vanishes. It means that the collective mode of pion exist until $T=1.15T_c$. The same plots for kaon are displayed in Fig. \ref{fig:coll_K}.  
The collective mode of kaon disappears at $T=1.15T_c$. This vanishing temperature of kaon pole is lower than that of pion.

\begin{figure}[htbp]
\begin{center}
\includegraphics[clip,width=100mm]{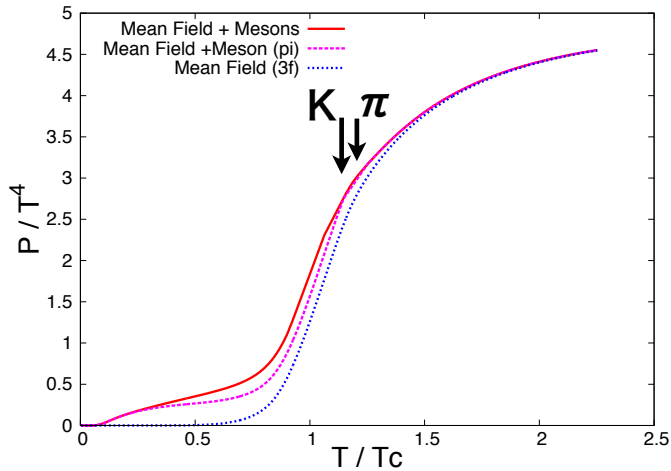}
\end{center}
\caption{
Pressure scaled by $T^4$ as a function of the temperature $T$ scaled by the pseudo critical temperature $T_c$.  
The solid red line is the sum of the contributions from the mean field and mesonic correlations (pion, kaon, sigma).  
The dotted pink line is for the mean field plus pionic correlation only and the dotted blue line is the pressure without mesonic correlations.  The melting temperatures of pions and kaons are indicated by arrows. }
\label{fig:EOS_3f_2}
\end{figure}

The vanishing points of pion pole and kaon pole are located on the places of the arrows in Fig. \ref{fig:EOS_3f_2}. 
It should be notable that pion and kaon still remain after the color-confinement is lost. Namely the quark-gluon excitations and meson excitations coexist in the transition region. The chiral symmetry is still not fully recovered at these temperatures.
There exist windows in the continuum of  quark-antiquark pair excitations in the $\omega-q$ plane where the spectrum of the isolated collective meson excitations reside. 
As the temperature increases further these windows are narrowed by the decrease of constituent quark masses and the collective meson spectra are absorbed in to the continua of individual quark-antiquark pair excitations. 

\section{Summary and concluding remarks}

We have studied the quark-hadron phase transition by using a three flavor PNJL model which contains the order parameters of both the chiral phase transition and the deconfining phase transition. 
In the mean field approximation the deconfining transition becomes a crossover transition in this model. 
%In this model thermal quark excitations are suppressed at low temperature by the Polyakov loop expressing the phase cancellation among quark distribution functions with three different color states. 
To include the mesonic thermal excitations, however, we need to go beyond the mean field approximation. 
We used the method of auxiliary fields to describe mesonic excitations;  
mesonic correlations are computed in the gaussian (one-loop) approximation for the fluctuation of the mesonic auxiliary fields, neglecting meson-meson interaction. 

We have extended our previous work with a two flavor case to a three flavor PNJL model incorporating nine pseudo scalar mesons and nine scalar mesons. % even though there remains   uncertainty of recognition of scalar mesons. 
Under the flavor SU(3) symmetry breaking, four kinds of pseudo scalar mesons($\pi $, $K$, $\eta $ and $\eta '$) and scalar mesons($\sigma $, $\kappa $, $a_0$, and $f_0$) appear with different masses and the equations of state are dominated by mesonic correlations, especially pions and kaons, at low temperatures.  
As temperature increases, the contribution from mesonic correlations decreases and the equation of state are dominated by quarks (and gluons given by hand). 
Mesonic excitations at low temperatures eventually melt and the degrees of freedom of thermal excitations change from hadrons to quarks. 
We have also studied the conditions for the existence of long-lived collective meson modes in pion and kaon channels 
and computed the melting temperatures of pions and kaons.   
We found that kaons melt first at around $T = 1.15T_c$ and pions melt at $ T = 1.2 T_c$ slightly above the melting point for kaons. 

In our calculation we only included the mesonic correlations in the pion, kaon, and $\sigma$ meson channels.  More massive mesons 
including $\sigma$ meson can decay into lighter mesons (pions) and it generates significant decay widths for these mesons.  Therefore, for a more complete treatment to include the effects of these higher mass mesonic correlations, we need to include the meson-meson interactions, taking into account the higher power terms in the expansion of the effective action written in terms of the mesonic auxiliary fields.  

In our model, the effect of confinement appears through the quark distribution function modified by the Polyakov loop 
expressing the phase interference among quark distribution functions with three different color states. 
In the vanishing Polyakov loop the modified quark distribution turn into the distribution of quark triads with the excitation energy three times larger than the single quark. 
The same modified quark distribution function appear also in the calculation of the temperature dependence of the collective modes, although their vacuum properties are the same as in the NJL model.   
This implies that the temperature dependence of the mesonic excitations show somewhat weaker than that calculated with the NJL model without confinement since the thermal excitations of quarks  are suppressed at low temperatures.
We note that the modified quark distribution for vanishing expectation value of the Polyakov loop is still not the same as the baryon distribution function as emphasized in our previous work\cite{Yamazaki:2012ux}.

This work has been done at zero baryon densities. However if we would try to extend our work to the finite chemical potential region, we must consider how to describe baryons as correlated three quark bound states. 

\section*{Acknowledgements}
We thank members of Komaba Nuclear Theory Group for their interests in this work.  
KY's work has been supported by the University of Tokyo Grants for Ph.D Research, 
Research Assistant for Creation of the Research Core in Physics,  
and School of Science Grants for PhD Students.   
TM's work has been supported by the Grant-in-Aid \# 25400247 of MEXT, Japan.  

\appendix
\def\thesection{Appendix \Alph{section}}
\renewcommand{\theequation}{A.\arabic{equation}}
\section{Explicit form of $\mathcal{F}(\omega , q)$} 

Here we show the explicit form of $\mathcal{F}(\omega, q)$ of pseudo scalar mesons. They can be written by using $\mathcal{F}_{uu}(\omega, q)$, $\mathcal{F}_{ss}(\omega, q)$  and $\mathcal{F}_{us}(\omega, q)$. Since we assume SU(2) isospin symmetry,  $\mathcal{F}_{ud}=\mathcal{F}_{dd}=\mathcal{F}_{uu}$.

\begin{eqnarray}
\mathcal{F}_{uu}(\omega, q)&=&\mathcal{F}_{uu}^{\bf scat}(\omega, q)+\mathcal{F}_{uu}^{\bf pair}(\omega, q) \\
&=&\int \frac{d^3p}{(2\pi )^3} \frac{1}{2E_u (p) 2E_u (p+q)} \nonumber \\
&\times &\Bigl( \frac{1}{\omega + E_u (p) - E_u (p+q)} 
-\frac{1}{\omega - E_u (p) + E_u (p+q)}\Bigr) \\
&\times &\bigl( f_{\Phi}(E_u(p))-f_{\Phi}(E_u(p+q))\bigr) \nonumber \\
&+&\int \frac{d^3p}{(2\pi )^3} \frac{1}{2E_u (p) 2E_u (p+q)} \nonumber \\
&\times &\Bigl( \frac{1}{\omega + E_u (p) + E_u (p+q)} 
-\frac{1}{\omega - E_u (p) - E_u (p+q)}\Bigr) \\
&\times &\bigl(1- f_{\Phi}(E_u(p))-f_{\Phi}(E_u(p+q))\bigr) \nonumber 
\end{eqnarray}

\begin{eqnarray}
\mathcal{F}_{ss}(\omega, q)&=&\mathcal{F}_{ss}^{\bf scat}(\omega, q)+\mathcal{F}_{ss}^{\bf pair}(\omega, q) \\
&=&\int \frac{d^3p}{(2\pi )^3} \frac{1}{2E_s (p) 2E_s (p+q)} \nonumber \\
&\times &\Bigl( \frac{1}{\omega + E_s (p) - E_s (p+q)} 
-\frac{1}{\omega - E_s (p) + E_s (p+q)}\Bigr) \\
&\times &\bigl( f_{\Phi}(E_s(p))-f_{\Phi}(E_s(p+q))\bigr) \nonumber \\
&+&\int \frac{d^3p}{(2\pi )^3} \frac{1}{2E_s (p) 2E_s (p+q)} \nonumber \\
&\times &\Bigl( \frac{1}{\omega + E_s (p) + E_s (p+q)} 
-\frac{1}{\omega - E_s (p) - E_s (p+q)}\Bigr) \\
&\times &\bigl(1- f_{\Phi}(E_s(p))-f_{\Phi}(E_s(p+q))\bigr) \nonumber 
\end{eqnarray}

\begin{eqnarray}
\mathcal{F}_{us}(\omega, q)&=&\mathcal{F}_{us}^{\bf scat}(\omega, q)+\mathcal{F}_{us}^{\bf pair}(\omega, q) \\
&=&\int \frac{d^3p}{(2\pi )^3} \frac{1}{2E_u (p) 2E_s (p+q)} \nonumber \\
&\times &\Bigl( \frac{1}{\omega + E_u (p) - E_s (p+q)} 
-\frac{1}{\omega - E_u (p) + E_s (p+q)}\Bigr) \\
&\times &\bigl( f_{\Phi}(E_u(p))-f_{\Phi}(E_s(p+q))\bigr) \nonumber \\
&+&\int \frac{d^3p}{(2\pi )^3} \frac{1}{2E_u (p) 2E_s (p+q)} \nonumber \\
&\times &\Bigl( \frac{1}{\omega + E_u (p) + E_s (p+q)} 
-\frac{1}{\omega - E_u (p) - E_s (p+q)}\Bigr) \\
&\times &\bigl(1- f_{\Phi}(E_u(p))-f_{\Phi}(E_s(p+q))\bigr) \nonumber 
\end{eqnarray}

Since pions are written as collective modes of u-quark, $\mathcal{F}_{\pi }$ is written by using only $\mathcal{F}_{uu}$;
\begin{eqnarray}
\mathcal{F}_{\pi}(\omega, q)=\mathcal{F}_{uu}(\omega, q)
\end{eqnarray}

For kaons, 
\begin{eqnarray}
\mathcal{F}_{K}(\omega, q)=\mathcal{F}_{us}(\omega, q) .
\end{eqnarray}

$\eta ^8$ and $\eta^0$ contains both $\bar{u}u$ and $\bar{s}s$.
For $\eta^8$ meson, 
\begin{eqnarray}
\mathcal{F}_{\eta^8}(\omega, q)=\mathcal{F}_{uu}(\omega, q)+2\mathcal{F}_{ss}(\omega, q)
\end{eqnarray}

For $\eta^0$ meson, 
\begin{eqnarray}
\mathcal{F}_{\eta^0}(\omega, q)=2\mathcal{F}_{uu}(\omega, q)+\mathcal{F}_{ss}(\omega, q)
\end{eqnarray}

\renewcommand{\theequation}{B.\arabic{equation}}
\section{Real part and imaginary part $\mathcal{F}(\omega , q)$} 

The function ${\cal F} (\omega \pm i \delta , q ) $ can be decomposed into real part ${\cal F}_1( \omega, q) $ and imaginary part 
$ {\cal F}_2 ( \omega, q ) $: 
\begin{eqnarray}
{\cal F} ( \omega \pm i \delta , q ) & = & {\cal F}_{1}( \omega, q) \pm i {\cal F}_2 (\omega, q ) 
%\nonumber \\
%{\cal F} ( \omega \pm i \delta , q ) =| {\cal F} (q, \omega \pm i \delta  ) | e^{\pm i \phi (q, \omega) } 
=  \sqrt{  {\cal F}_1( \omega, q )^2 +  {\cal F}_2 ( \omega, q )^2 } e^{\pm i \phi ( \omega, q) } \ \ \ 
\end{eqnarray}
%with the modulus and the argument given by
%\begin{eqnarray}
%& & | {\cal F} (q, \omega \pm i \delta  ) | = \sqrt{  {\cal F}^1(q, \omega)^2 +  {\cal F}^2 (q, \omega )^2 } \\
%\end{eqnarray}
where the argument $\phi$ is given by
\begin{equation}
 \phi (\omega, q) =  \tan^{-1} \frac{{\cal F}_2(\omega, q)}{{\cal F}_1(\omega, q)} ~~. \label{argumentofF}
\end{equation}
Ancillary to that, $\mathcal{M}(\omega \pm i \delta , q )$ has also an imaginary part. 
The real part and imaginary part of the function ${\cal F}$ are further decomposed into two parts: 
the scattering term and the pair excitation term.
The two components of the real part are given by the principal part integrals: 
\begin{eqnarray}
{\cal F}_{ij, 1}^{\rm scatt.} (\omega, q) %& = & Re F(q, \omega + i \delta ) \nonumber \\
& = & {\cal P} \int \frac{d ^3 p}{(2\pi)^3} \frac{1}{2 E_i(p) 2 E_j(p+q)} \\
& & \qquad \times \left( \frac{1}{\omega + E_i(p) - E_j(p+q)} - \frac{1}{\omega - E_i(p) + E_j(p+q)} \right) 
\nonumber \\
& &  \qquad  \qquad \qquad \qquad  \times \left( f_\Phi (E_i(p)) - f_\Phi (E_j(p+q) ) \right) \nonumber \\ 
{\cal F}_{ij, 1}^{\rm pair} ( \omega, q) %& = & Re F(q, \omega + i \delta ) \nonumber \\
& = & {\cal P} \int \frac{d ^3 p}{(2\pi)^3} \frac{1}{2 E_i(p) 2 E_j(p+q)} \\
%& &  \left. \qquad \qquad \qquad \qquad +  
& & \qquad \times \left( \frac{1}{\omega + E_i(p) + E_j(p+q)} - \frac{1}{\omega - E_i(p) - E_j(p+q)} \right)
\nonumber \\
& &  \qquad  \qquad \qquad \qquad \qquad  \times  
\left( 1 - f_\Phi (E_i(p)) - f_\Phi (E_j(p+q) ) \right) \nonumber 
% \right] \nonumber \\
\end{eqnarray}
while the two components of the imaginary part contains the energy conserving $\delta$-functions:
\begin{eqnarray}
{\cal F}_{ij, 2}^{\rm scatt.} (\omega, q) %& = &  Im F(q, \omega + i \delta ) \nonumber \\ 
& = &  - \pi  \int \frac{d ^3 p}{(2\pi)^3} \frac{1}{2 E_i(p) 2 E_j(p+q)}
\left( f_\Phi (E_i(p)) - f_\Phi (E_j(p+q) ) \right) \nonumber \\
& &  \times \left( \delta (\omega + E_i(p) - E_j(p+q)) - \delta ( \omega - E_i(p) + E_j(p+q) ) \right)  \\ 
{\cal F}_{ij, 2}^{\rm pair} (\omega, q) 
& = &  - \pi  \int \frac{d ^3 p}{(2\pi)^3} \frac{1}{2 E_i(p) 2 E_j(p+q)} %\left. \qquad \qquad \qquad \qquad +  
\left( 1 - f_\Phi (E_i(p) ) - f_\Phi (E_j(p+q) ) \right) \nonumber \\
%\left( \frac{1}{\omega + E_i(p) + E_j(p+q)} - \frac{1}{\omega - E_i(p) - E_j(p+q)} \right) 
& & \left( \delta (\omega + E_i(p) + E_j(p+q)) - \delta ( \omega - E_i(p) - E_j(p+q) ) \right) 
% \right] \nonumber \\
\end{eqnarray}
It is evident that the scattering term has non-zero imaginary term in the space-like energy-momentum region 
($\bq^2 > \omega^2$), while the pair creation/annihilation term is non-vanishing only in the time-like region
($\bq^2 < \omega^2$).  
It is important to note that non-collective mesonic correlation arises only from non-vanishing 
imaginary part of ${\cal F} (q, \omega)$.

%%%%%%%%%%%%%%%%%%%%%%%%%%%%%%%%%%%

%%%%%%%%%%%%%%%%%%%%%%%%%%%%%%%%%%%%%%%%

\end{document}